\documentclass[prb, floatfix,longbibliography,reprint, twocolumn,superscriptaddress]{revtex4-2}
\usepackage[utf8]{inputenc}
\usepackage[T1]{fontenc}
\usepackage{amsmath}
\usepackage{graphicx}
\usepackage{xcolor}
\usepackage[export]{adjustbox}
\usepackage{bm}
\usepackage{setspace}
\usepackage{comment}
\usepackage{bold-extra}
\usepackage[shortcuts]{extdash}

\newcommand{\R}{\boldsymbol{R}}

\newcommand{\figurewidth}{7.5cm}
\renewcommand{\eqref}[1]{Eq.~(\ref{#1})}

\graphicspath{{figures/}}

\begin{document}

\title{Estimating melting curves for Cu and Al from simulations at a single state point}

\date{\today}
\author{Laura Friedeheim}
\affiliation{``Glass and Time'', IMFUFA, Dept. of Science and Environment, Roskilde
 University, P. O. Box 260, DK-4000 Roskilde, Denmark}
\author{Felix Hummel}
\affiliation{Institute for Theoretical Physics, TU Wien,
  Wiedner Hauptstrasse 8--10, A-1040 Vienna, Austria}
\author{Jeppe C. Dyre}
\affiliation{``Glass and Time'', IMFUFA, Dept. of Science and Environment, Roskilde
 University, P. O. Box 260, DK-4000 Roskilde, Denmark}
\author{Nicholas P. Bailey}
\affiliation{``Glass and Time'', IMFUFA, Dept. of Science and Environment, Roskilde
 University, P. O. Box 260, DK-4000 Roskilde, Denmark}

\begin{abstract}
Determining the melting curves of materials up to high pressures has long been a challenge experimentally and theoretically. 
A large class of materials, including most metals, has been shown to exhibit hidden scale invariance, an approximate scale invariance of the potential-energy landscape that is not obvious from the Hamiltonian. For these materials the isomorph theory allows the identification of curves in the phase diagram along which structural and dynamical properties are invariant to a good approximation when expressed in appropriately scaled form. These curves, the isomorphs, can also be used as the basis for constructing accurate melting curves from simulations at a single state point [U. R. Pedersen \textit{et al.}, Nat. Comm. \textbf{7}, 12386 (2016)]. In this work we apply this method to the  metals Cu simulated using the effective medium theory and Al simulated using density functional theory (DFT). For Cu the method works very well and is validated using two-phase melting point simulations. For Al there are likewise good isomorphs, and the method generates the melting curve accurately as compared to previous experimental and DFT results. In support of a recent suggestion of Hong and van de Walle [Phys. Rev. B \textbf{100}, 140102 (2019)], we finally suggest that the tendency for the density-scaling exponent $\gamma$ to decrease with increasing density in metals implies that metals in general will undergo re-entrant melting, i.e., have a maximum of melting temperature as a function of pressure.
\end{abstract}

\parskip 0ex
\parskip 1ex

\maketitle

\section{Introduction}

Melting is the physical process where a substance changes phase from solid to liquid.\cite{Chandler:1987,Atkins:1990,Glicksman:2011} For a pure one-component substance at a given pressure, the melting temperature $T_m$ is the temperature at which both solid and liquid phase can coexist in equilibrium. In general, the melting temperature depends on the pressure and the term ``melting curve'' refers to the functional dependence of the melting temperature on the pressure in the phase diagram. 

Melting of metals --- especially at high temperatures and pressures --- is of particular interest for a variety of disciplines from material science to geophysics and planetary science. For example, understanding the conditions in the core of the earth has long been an intriguing endeavor, evidenced by the many papers on high-pressure melting of iron and iron-rich alloys alone.\cite{Birch:1952,Williams/others:1987,Stixrude/Wasserman/Cohen:1997,Alfe/others:2007,Tateno/others:2010,Anzellini/others:2013,Wilson/others:2023} Determining high-pressure melting curves in metals in general is an active field, both on the experimental side \cite{Errandonea/others:2001,Hanstrom/Lazor:2000,Errandonea:2010,Errandonea:2013} and from theory and simulations.\cite{Hieu/Ha:2013,Hieu:2014,Zhang/others:2014,Gubin/others:2015} 

Predictions of melting curves have a long history, starting with the first attempt made by Lindemann \cite{Lindemann:1910} in 1910. The Lindemann melting criterion, which in its well-known form is actually an extension to Lindemann's original work made several years later by Gilvarry,\cite{Gilvarry:1956} states that melting approximately occurs when the root-mean-square amplitude of the thermal vibration exceeds the threshold value of 10\% of the nearest-neighbor distance. 

In simulations, the melting temperature can be determined similarly to experiments by observing the phase transition directly, such as the fast heating (Z-method) where a small slab of initially solid particles is heated until it melts.\cite{Belonoshko/others:2006,Raty/others:2007} Without particular nucleation sites such as surface defects that are available for melting in nature, however, this procedure tends to lead to a meta-stable, super-heated crystal phase prior to melting and thus overestimates the melting temperature.\cite{Belonoshko/others:2013} 
This is sometimes counteracted by combining the temperature found from melting with the temperature found from observing the phase transition when freezing the same system. The reversed process results in an underestimated temperature following from the absence of crystallization seeds from which crystallization can start more easilly.

Accurate estimates of the melting temperature can be given by coexistence methods, provided the system is large enough.\cite{Morris/others:1994,Ogitsu/others:2003,Alfe:2005,Hong/Walle:2013} Here, solid and liquid phases are brought in direct contact with each other. The interface between the phases is monitored, and the coexistence state point is inferred from the interface position when the system has stabilized. The interface-pinning method \cite{Pedersen:2013} is a development of the coexistence method, where the interface is preserved in equilibrium by an extra term in the Hamiltonian. An advantage of this is that the system can be stabilized in a much smaller cell and thus requires less computationally expensive simulations while maintaining accuracy.

All these techniques only determine a single point $(p,T_m)$ on the melting curve at a time. Tracing out all points on the melting curve in this way using {\em ab initio} simulations can be prohibitively costly computationally. Thus, predictions of the melting curve usually involve some kind of theoretical interpolation or extrapolation to make a curve out of only a few points. Reflecting the traditional perception of melting, i.e., higher pressure gives a higher melting temperature, these procedures usually lead to predictions where the temperature monotonically rises with increasing pressure. This includes, for example, the widely accepted Lindemann and Gr\"uneisen laws \cite{Lindemann:1910,Gilvarry:1966,Ross:1969} as well as the Simon--Glatzel equation.\cite{Simon/Glatzel:1929} 

Recent \emph{ab initio} simulations have shown that a wide class of materials, including metals, approximately obey a hidden scale invariance.\cite{Hummel/others:2015} This means that curves exist in the phase diagram called isomorphs along which structure and dynamics are approximately invariant in properly chosen units.\cite{paper4,Schroder/Dyre:2014,Dyre:2014} The isomorphs are almost parallel to the melting curve, and an isomorph starting from a state point at coexistence can be regarded as a zeroth-order approximation to the melting curve in the face of hidden scale invariance. Indeed, if the property of structural invariance held for two-phase configurations, the melting curve would be exactly an isomorph. Building on this, Ref. \onlinecite{Pedersen/others:2016a} gives a prescription of how the pressure-temperature melting curve, as well as the freezing and melting densities, can be recovered over a wide range of pressures by a first-order Taylor expansion from the two reference isomorphs of the solid and the liquid constructed at a single coexistence point. The method involves sampling relatively few configurations from one reference-point simulations and scaling them to the densities where an estimate for the melting pressure and temperature is desired. Instead of performing molecular dynamics or Monte Carlo simulations it suffices to evaluate the total energy and the virial of the scaled configurations to obtain both, an estimate for the melting pressure and the melting temperature at the corresponding density. This involves much less computational effort than performing simulations at the respective densities and temperatures. For example, if the configurations are sampled every 100 simulation steps, the computational work to calculate the melting curve is of order 100 times smaller per point on the melting curve in the region where interpolation or extrapolation is used.
This method has already been validated for the Lennard--Jones system and for realistic potentials for noble elements.\cite{Pedersen/others:2016a,Singh/Dyre/Pedersen:2021}
The present paper investigates this method for the melting of metals based on \textit{ab initio} and effective medium theory simulations and demonstrates its computational advantage.

The proposed method is limited to regions where hidden scale invariance holds qualitatively.
Although this may be a pressure range covering hundreds of GPa this method cannot
predict structural transitions, such as fcc-bcc transitions,\cite{Baty/Burakovsky/Errandonea:2021,Smirnov:2021} Peierls-type transitions\cite{Gehring/Gehring:1975,Koppel/Yarkony/Barentzen:2011,Raty/others:2007}
or metal-insulator transitions.%
\cite{MetalToNonmet2010}
In the case of bcc-fcc transitions of metals hidden scale invariance is still useful to individually
trace the two coexistence curves of the liquid and either of the solid phases.
The liquid-bcc-fcc triple point needs to be found in a separate calculation
and marks the transition from one interpolation region to the next.
In case of metal-insulator phase transitions hidden scale invariance is qualitatively
absent in the insulating phase and the proposed melting curve prediction method
is only useful for metallic phases.

An interesting by-product of the analysis in terms of isomorphs is that re-entrant melting may likely be a general feature of metals. Re-entrant melting is where the melting curve reaches a maximum temperature at some high pressure before decreasing at even higher pressure; at a fixed temperature below the maximum one can go from liquid to crystal and back to liquid again only by increasing pressure. This is supported by a recent density functional theory (DFT) study of melting curves of metallic elements at high pressure by Hong and van de Walle suggesting that re-entrant melting is much more widespread than previously realized.\cite{Hong/Walle:2019}


The structure of the paper is as follows. In Section~\ref{sec:isomorphs} we present an overview of isomorph theory, including the method used to generate isomorphs and the method for determining the melting curve from two isomorphs (one liquid and one crystal), starting at a known point on the melting curve. Sec.~\ref{sec:melting_Cu} presents the melting curve method to Cu simulated using effective medium theory (EMT). Sec~\ref{sec:isomorph_invariance_DFT} investigates the degree of isomorph invariance of structural and dynamical quantities for Al along the computed isomorphs using DFT and 
presents our results for the melting curve of Al. A discussion of the implications for re-entrant melting is given in Sec.~\ref{sec:reentrant_melting}.

\section{Hidden scale invariance and isomorphs}\label{sec:isomorphs}

Isomorph theory has been developed over a series of papers.\cite{paper1,paper2,paper3,paper4,paper5} An updated, generic version of the theory can be found in Ref.~\onlinecite{Schroder/Dyre:2014}, and reviews in Refs.\ \onlinecite{Dyre:2014,Dyre:2016,Dyre:2018}. The following introduces briefly the concepts of isomorph theory that are relevant for the present paper. A key concept is {\em hidden scale invariance}, which refers to an underlying approximate symmetry that makes the phase diagrams of materials possessing this symmetry particularly simple. ``Simplicity'' in this sense of the word is referred to as R-simplicity.\cite{Schroder/Dyre:2014} Hidden scale invariance is defined by the property that the potential energies of same-density configurations maintain their ordering under uniform volumetric scaling. This condition can be expressed\cite{Schroder/Dyre:2014} as 
\begin{equation}
    U(\R_a) < U(\R_b) \, \implies \, U(\lambda \R_a) < U(\lambda \R_b),
    \label{eq:ordering}
\end{equation}
where $U(\R_i)$ is the potential energy of the configuration $\R_i$ (i.e., all particle coordinates) and $\lambda$ is a scaling parameter. This condition is obeyed to a good approximation in the condensed part of the phase diagram of various systems, including both solid and liquid phases of real as well as model systems like the Lennard--Jones and Yukawa systems. Equation (\ref{eq:ordering}) is general: it does not assume equilibrium configurations or, if these are in fact equilibrated, does not assume the same temperature. It has been shown in simulations that hidden scale invariance is often spoiled by directional interactions as well as by competing length scales when more than one kind of interaction is present.\cite{paper2} Thus, while most metals and van der Waals bonded molecular systems are expected to be R-simple, hydrogen and covalently-bonded systems are not. Ionic and dipolar systems constitute an interesting in-between case.\cite{Dyre:2014,Knudsen/Niss/Bailey:2021}

\subsection{Strong virial potential-energy correlations}

Systems that for most of their configurations obey the condition of \eqref{eq:ordering} have previously also been referred to as \emph{strongly correlating}.\cite{paper1,paper2} Recall that the virial $W$ at a given state point is the contribution from interactions to the pressure via $PV=Nk_BT+W$; $W$ is an extensive quantity of dimension energy. The strong correlation refers to the instantaneous equilibrium fluctuations of the potential energy $U$ and $W$,

\begin{equation}
    \Delta W \cong \gamma \Delta U \label{eq:gamma_correlation},
\end{equation}
where $\Delta$ indicates instantaneous deviations from the canonical constant-volume ($NVT$) ensemble average. A system is considered strongly correlating in the isomorph sense if $R~>~0.9$, where $R$ is the (Pearson) correlation coefficient
\begin{equation}
    R=\frac{\langle \Delta W \Delta U \rangle}{\sqrt{\langle (\Delta W)^2\rangle \langle(\Delta U)^2\rangle}},
    \label{eq:R}
\end{equation}

with the angle brackets denoting $NVT$ expectation values. The proportionality factor $\gamma$ between virial and potential-energy fluctuations is given as the generally state-point dependent linear-regression slope (in which $\rho$ is the particle number density)

\begin{equation}
    \gamma(\rho, T) = \frac{\langle \Delta W \Delta U \rangle}{\langle (\Delta U)^2 \rangle}.
    \label{eq:gamma}
\end{equation}

The factor $\gamma$ is called the \textit{density-scaling exponent}.\cite{paper1} This quantity may be determined by application of the generally valid equality, derived in Ref.~\onlinecite{paper4},

\begin{equation}
    \gamma(\rho, T) = \left( \frac{\partial \ln T}{\partial \ln \rho}  \right)_{S_{\rm ex}},
    \label{eq:gamma_diff}
\end{equation}

in which $S_{\rm ex}$ is the excess entropy referring to the deviation from the ideal-gas entropy at the same temperature and density. Thus, as a thermodynamic quantity, $\gamma$ gives the slope in the logarithmic density-temperature phase diagram of curves of constant excess entropy. The equality of \eqref{eq:gamma} and \eqref{eq:gamma_diff} is a general statistical-mechanical identity valid for any system.\cite{paper4}

Curves of constant excess entropy exist for any system; for an R-simple (strongly correlating) system such curves are called \textit{isomorphs}. Thus systems only have isomorphs in the part of their phase diagram where they are strongly correlating or, equivalently, where most of the physically relevant configurations obey \eqref{eq:ordering}. Along isomorphs, structure and dynamics are invariant to a good approximation, and this fact effectively reduces the phase diagram by one dimension. To observe the invariance, quantities must be rescaled into appropriate dimensionless form using so-called reduced units. For example, lengths at different densities can only be directly compared after dividing by the average interparticle spacing $\propto\rho^{-1/3}$. Likewise, times are given in multiples of the time a particle at thermal velocity needs to pass the interparticle spacing (apart from a numerical factor of order unity), $\rho^{-1/3}\sqrt{m/k_B T}$, and energies are scaled by the thermal energy $k_B T$.\cite{paper4}

\begin{figure}
  \begin{center}
    \includegraphics[width=\columnwidth]{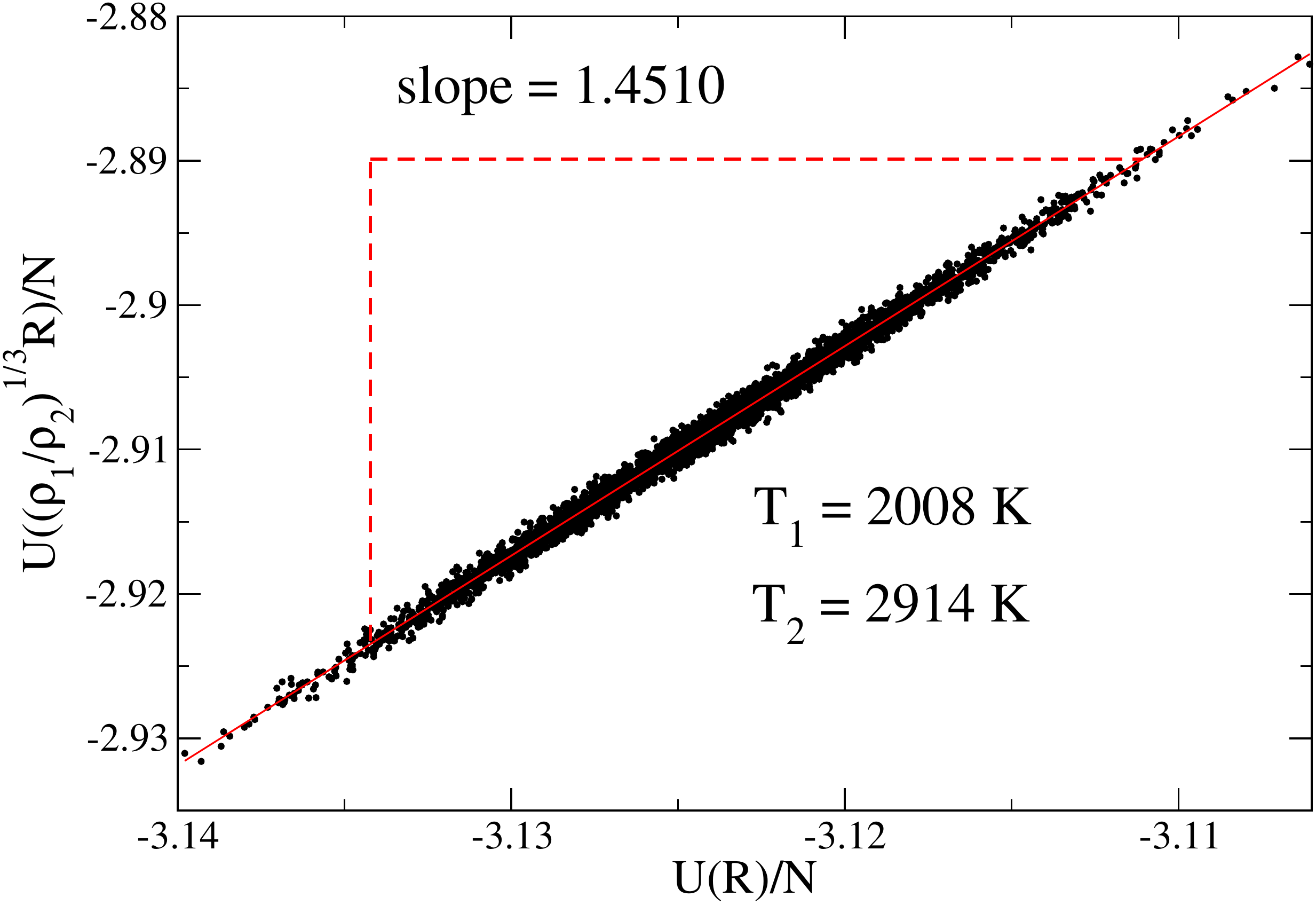}
    \end{center}
    \caption{Determining the temperature $T_2$ at a scaled density $\rho_2$ from an EMT simulation of 2048 atoms of liquid Cu at $T_1=2008$~K, $\rho_1=0.0831$~\AA${}^{-3}$, corresponding to $P=16.0$~GPa. The unscaled potential energies are taken from configurations of the initial simulation. The same configurations are then uniformly scaled to the target density, here $\rho_2 = 1.1312 \rho_1$. The direct isomorph check finds the temperature $T_2$ for the given density $\rho_2$ of a state point that is on the same isomorph as the initial state point $(\rho_1, T_1)$. Plotting the potential energies of scaled versus unscaled configurations results in a scatter plot where the slope of the best fit line is $T_2/T_1$ (see Eq.~(\ref{eq:DIC})), resulting in the value 2914 K for $T_2$.}
    \label{fig:DIC}
\end{figure}

\subsection{Direct isomorph check}\label{sec:DIC}

Isomorphs can be traced out in a step-wise fashion
, where $\gamma$ can be found from fluctuations in simulations at each state point via \eqref{eq:gamma} and numerically integrating \eqref{eq:gamma_diff} by the Euler or the Runge-Kutta algorithm, changing density in small increments. This is, however, not practical for computationally intensive simulations like those based on \emph{ab initio} methods, as this method requires simulating at every step. Instead, the present work uses the so-called direct isomorph check (DIC).

The DIC was introduced in Ref. \onlinecite{paper4} and works as follows. For any two state points on the same isomorph one can make a one-to-one correspondence between their respective microscopic configurations $\boldsymbol{R}_i = (\boldsymbol{r}_1^{(i)},... \boldsymbol{r}_N^{(i)})$, specifically between those that have the same reduced coordinates, $\rho_1^{1/3} \boldsymbol{R}_1 = \rho_2^{1/3} \boldsymbol{R}_2$, i.e., can be scaled uniformly into one another. Then, according to the isomorph theory,\cite{paper4} the corresponding configurations have almost identical configurational $NVT$ canonical probabilities,
\begin{equation}
     \exp \left( - \frac{U ( \boldsymbol{R}_1 )}{ k_B T_1} \right) \cong C_{12} \exp \left( - \frac{U ( \boldsymbol{R}_2 )}{ k_B T_2} \right)\,,
     \label{eq:boltzmann}
\end{equation}
where $C_{12}$ is a constant specific to this pair of state points. Using this, \eqref{eq:boltzmann} expressed in terms of fluctuations around the respective means becomes
\begin{equation}
    \Delta U ( \boldsymbol{R}_2 ) = \Delta U \left( \left(\frac{\rho_1}{\rho_2} \right)^{1/3}\boldsymbol{R}_1 \right) \cong \frac{T_2}{T_1} \Delta U ( \boldsymbol{R}_1 )\,.
    \label{eq:DIC}
\end{equation}
From this the meaning of the slope when plotting the energies as shown in Fig.~\ref{fig:DIC} becomes clear: plotting the potential energies from configurations at a state point $(\rho_1, T_1)$ against those determined from the same configurations scaled to another density $\rho_2$, results in a scatter plot whose slope is $T_2/T_1$, thus allowing $T_2$ to be determined.\cite{paper4,Schroder/Dyre:2014} The same initial configurations can be scaled to several different densities, so that an isomorph can be mapped out from simulations at only one state point (although the quality of the linear fit generally degrades as the density change increases).

Reasonable estimates of thermal averages of quantities -- such as the potential energy, virial, pressure -- at the scaled densities can be made by averaging over the scaled configurations. Indeed, since these quantities are not invariant along isomorphs in reduced units, non-trivial thermodynamic data can be obtained this way. Even fluctuation-based quantities like $\gamma$ can be estimated from the scaled configurations. This is the key to being able to determine thermodynamic quantities along isomorphs without doing  simulations except at the reference state point. Calculating thermodynamic quantities such as virial and potential energy, can still be costly, especially using {\em ab initio} methods, but two orders of magnitude less costly than doing simulations at the scaled densities if configurations are sampled every 100 time steps. It must be noted that the scaled configurations cannot tell us how invariant structure and dynamics actually are at any of the scaled densities. To assess the degree of invariance conducting simulations at the scaled densities is unavoidable.

\subsection{Relevance of isomorphs to the melting curve}

Paper IV of the initial series of papers developing the isomorph theory \cite{paper4} argues that the melting curve must be parallel to liquid and solid isomorphs since an isomorph crossing the melting curve contradicts the isomorph invariance of structure. Obviously, if structure is invariant, the system cannot at the same time undergo a phase transition. This argument assumes a very strong structural invariance applying to configurations containing a mixture of two phases of different densities.

For realistic systems, the preservation of ordering of \eqref{eq:ordering} is satisfied for most, but not all, pairs of physically relevant configurations. Isomorph theory is only exact in systems with potential-energy functions which are Euler-homogeneous of degree $n$ (plus a constant). In this case the correlation coefficient and the density-scaling exponent are $R=1$ and $\gamma=n/3$, respectively. Inverse power law (IPL) potential systems\cite{paper3} are an example of this -- there the isomorph starting from a point at melting simply follows the melting curve. 

In more realistic systems the strong isomorph requirement that structure and dynamics be invariant also for two-phase configurations must be relaxed to the weaker condition that isomorph invariance applies for single-phase configurations. This is due to the fact that the density-scaling exponent generally depends on density, so identical density scaling of liquid and crystal phases will scale their potential energy surfaces by different amounts. This means that the melting and freezing lines are not exact isomorphs, though still close to isomorphs. Isomorphs can be considered as excellent zeroth-order approximations of the melting and freezing lines. Indeed, several phenomenological melting rules, including the Lindemann melting criterion, can be understood as consequences of the melting curve being close to an isomorph.\cite{paper4}

\begin{figure}[t]
    \flushleft
    \includegraphics[width=\figurewidth]{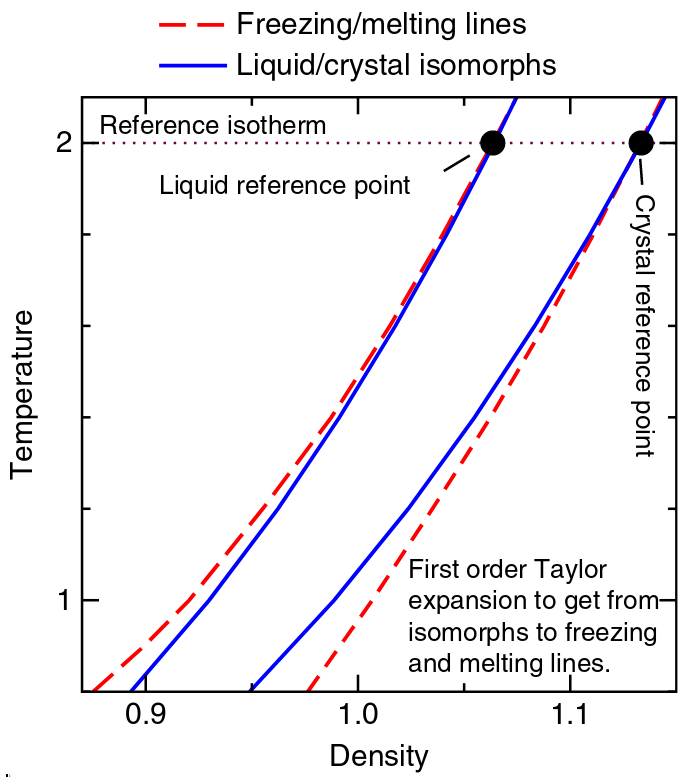}
    \caption{Figure taken from Ref.~\onlinecite{Pedersen/others:2016a} illustrating the idea behind Eq.~(\ref{eq:meltingP}) (for the Lennard--Jones system). It shows isomorphs (blue) generated from simulations at reference points (black dots) on the freezing/melting curve. The method then gives a prescription to interpolate the freezing/melting curve (red dashed) beyond the reference point as a first-order expansion from the reference isomorphs. The present work, in contrast to the figure, takes a low reference temperature and generates melting curves towards higher temperatures.}
    \label{fig:LJmeltingIdea}
\end{figure}

The method we use for constructing accurate melting curves was described in Ref.~ \onlinecite{Pedersen/others:2016a} using the standard single-component Lennard--Jones system as an example. The basic idea is illustrated in Figure~\ref{fig:LJmeltingIdea}. Two isomorphs, one liquid and one solid, are generated from a state point at coexistence. From quantities along these ``reference isomorphs'', the melting pressure as well as the freezing and melting densities can be calculated. The resulting expression for the melting pressure is
\begin{align}
  \nonumber
    &P_m(T) = \Bigg[ \left( U_s^I - \frac{T}{T^0}U_s^0 \right) - \left( U_l^I - \frac{T}{T^0}U_l^0 \right) \label{eq:meltingP}
    \\
    & + N k_B T \ln{\left(\Tilde{\rho}^I_s/\Tilde{\rho}^I_l\right)} + \frac{T}{T^0}\left( W_l^0 - W_s^0 \right) \Bigg] \frac{1}{\left(V_l^I - V_s^I\right)}\,.
\end{align}
Here $\Tilde{\rho}^I_{l,s}=\rho^I_{l,s}/\rho^0_{l,s}$ are the densities along the liquid (l) and the solid (s) isomorph relative to their reference values, respectively, and $U^I_{l,s}$, $V^I_{l,s}$, and $W^I_{l,s}$ are the potential energies, volumes, and virials along the liquid and solid isomorphs at the temperature $T$. The temperature dependence of $U$, $V$, and $\tilde \rho$ has been suppressed for compactness in \eqref{eq:meltingP}. Knowing the deviation of the pressure between the melting curve and the reference isomorph, the liquid density at freezing (f) and solid density at melting (m) can be found using
\begin{align}
    \rho_{f,m} (T) \cong \rho_{l,s} (T) \left( 1 + \frac{P_m (T) - P_{l,s}^I}{K_{T \, l,s}^I} \right),
    \label{eq:meltingrho}
\end{align}
respectively,
where $K_{T \, l,s}^I$ are the isothermal bulk moduli along the liquid and the solid isomorphs, respectively. In an equilibrium simulation this quantity is calculated from the following fluctuation formula:\cite{Allen/Tildesley:1987}
\begin{equation}\label{eq:K_T_formula}
    K_T = \frac{ 1 }{V} \left( N k_B T + W + X - \frac{\langle(\Delta W)^2\rangle}{k_B T } \right).
\end{equation}

Here is $X$ is the average of the so-called hypervirial, which is defined for individual configurations by $\partial W/\partial \ln \rho$ for a uniform scaling of all coordinates.\cite{Allen/Tildesley:1987} Together, Eqs. (\ref{eq:meltingP}) and (\ref{eq:meltingrho}) give a prediction for the melting curve in the pressure-temperature diagram, as well as for the melting and freezing curves in the density-temperature diagram, using quantities along the isomorphs. Moreover, as described at the end of Sec.~\ref{sec:DIC}, these quantities can be estimated without doing additional simulations along the isomorphs. The hypervirial requires additional calculations at nearby densities to calculate the numerical derivative with respect to the density, however.

In Eqs. (\ref{eq:meltingP}) and (\ref{eq:meltingrho})  the input quantities along the isomorphs are to be evaluated at the same temperature. In Ref.~\onlinecite{Pedersen/others:2016a} the method was applied to the Lennard--Jones system where analytical expressions for thermodynamic quantities along the isomorphs exist.\cite{paper5} This is not the case for the models employed in this work. Additionally, the DIC finds temperature (and other thermodynamic quantities) for a specific density. Thus, the data along the two isomorphs are not readily available as a function of temperature, which the method requires. This is handled by fitting the temperature dependence of all DIC data with polynomials. This procedure is a possible source of error, and different orders of polynomials may be needed for the different quantities. In particular, we find that the potential-energy contribution to \eqref{eq:meltingP} poses the largest source of error emerging from small differences between large numbers. The other two terms remain fairly constant. Fourth-order polynomials were used to fit  the potential energy as a function of temperature, while third-order polynomials were used for the other quantities.

\section{Isomorphs and melting curve of EMT Copper}\label{sec:melting_Cu}

\subsection{Effective medium theory many-body potentials}

To investigate the melting curve and its associated isomorphs in realistic models of metals, we have first carried out simulations using the Roskilde University Molecular Dynamics (RUMD) \cite{Bailey/others:2017} code for molecular dynamics simulations on GPUs using the effective medium theory (EMT) potential.\cite{Jacobsen/others:1987,Jacobsen/others:1996} The EMT potential is a semi-empirical interatomic potential that aims to combine the accuracy of DFT with the computational efficiency and transparency of a simple pair potential. The basis of the EMT potential is the ansatz that the energy of an atom inserted into an inhomogeneous electronic medium is a function of the electron density at the atom's location, or more generally averaged over the volume of the atom. This allows the calculation of atoms in inhomogeneous environments to be determined from the immersion energy of an atom in a uniform electron gas (atom-in-jellium model). It has therefore been described as a ``local density approximation for atoms''. The validity of this ansatz was backed up with extensive DFT calculations in the 1980's.\cite{Puska/Nieminen/Manninen:1981,Puska/Nieminen:1991} From these results explicit parameterizations of the embedding energy for many elements have been determined. To make an interatomic potential based on the ansatz, the further assumption is made that the effective electron density into which a given atom is embedded may be calculated as a superposition of the electron density tails from the neighboring atoms.\cite{Jacobsen/others:1987,Jacobsen/others:1996} Corrections to this picture may be implemented using a simple pair potential, and the resulting interatomic potential is efficient, using simple functional forms (mostly exponentials) and relatively few parameters.

This is in contrast to well-known Embedded-Atom Method (EAM) potentials \cite{Murray/Baskes:1983,Mishin/others:2001,Mitev/others:2006,Mendelev:2009} whose overall structure and many-body nature is similar, but which tend to be based on heavy fitting to, e.g.\ experimental structural data, and which do not lend themselves as much to obtaining physical insight. One feature of the EMT potential is that it gives reasonably accurate values for the isomorph-theory density-scaling parameter $\gamma$ when compared to DFT calculations, a quantity to which the potentials were not fitted in any way.\cite{Friedeheim/others:2019} A practical advantage of EMT's simplicity compared to EAM is its ease of implementation in RUMD, but it also allows in principle the possibility of an analytical investigation of the dependence of $\gamma$ on the potential parameters and on the density.

\begin{figure}[t!]
    \centering
    \includegraphics[width=0.75\linewidth]{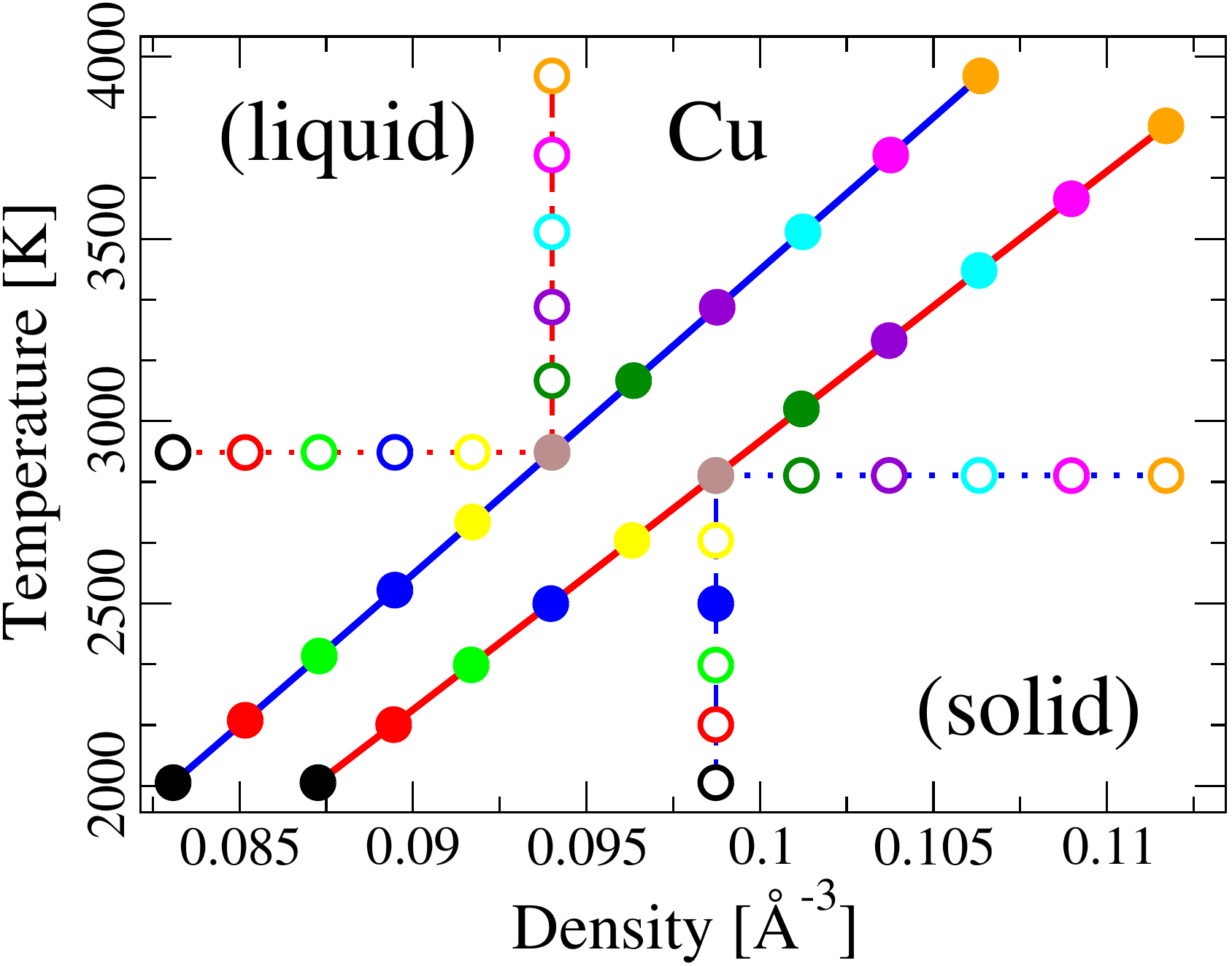}
    \caption{Relevant part of the Cu phase diagram marking the points of interest. The solid lines denote isomorphs generated from the reference state points given by the black circles (lower left corner). Additional simulations were carried out at the state points marked by open circles. The points connected by the dotted/dashed line are on the same isotherm/isochore. All iso-curves are given in blue for the liquid phase and in red for the solid phase. The colors of the circles correspond to the colors used in Figs.~\ref{fig:Cu_rdf+msd_solid} and \ref{fig:Cu_rdf+msd_liquid} for structure and dynamics data at the respective state points.
    }
    \label{fig:Cu_phase_diagram}
\end{figure}

\begin{figure}[t!]
    \centering
    \includegraphics[width=0.7\linewidth]{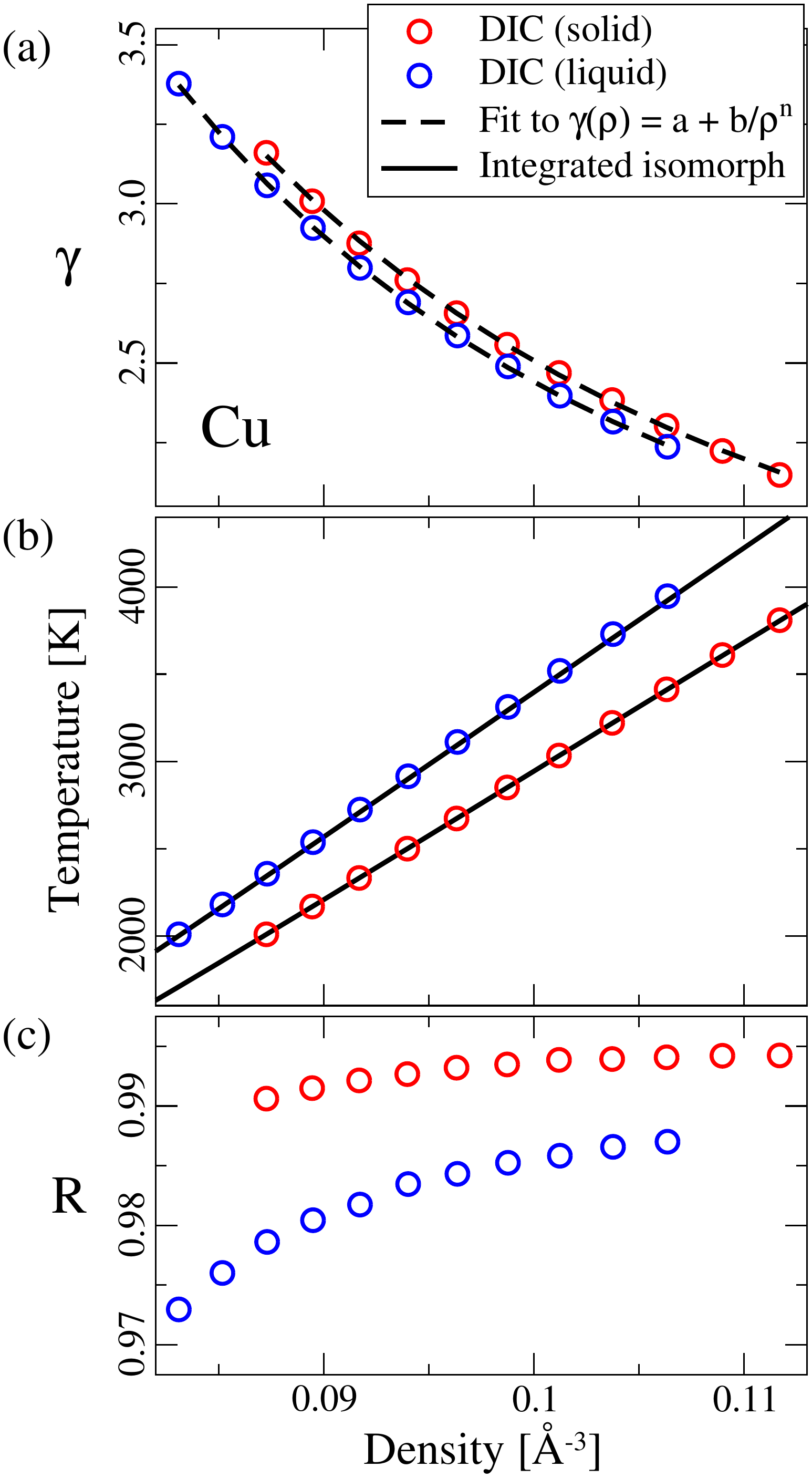}
    \caption{DIC self-consistency check for EMT-Cu. Open circles represent data for the isomorphs from the DIC, red for the solid, and blue for the liquid. The dashed line in the top panel was obtained from fitting to the DIC points. The parameter for the fitted expression for $\gamma(\rho)$ are: $a=1.349, b=0.0006528, n=3.249$ for the solid and $a=1.409, b=0.0003428, n=3.478$ for the liquid isomorph. The resulting integrated isomorph is given by the solid line in the middle panel. The bottom panel shows the correlation coefficients along both isomorphs.
    %
    }
    \label{fig:Cu_DIC_consistency}
\end{figure}

\begin{figure}
\centering
   \includegraphics[width=\linewidth]{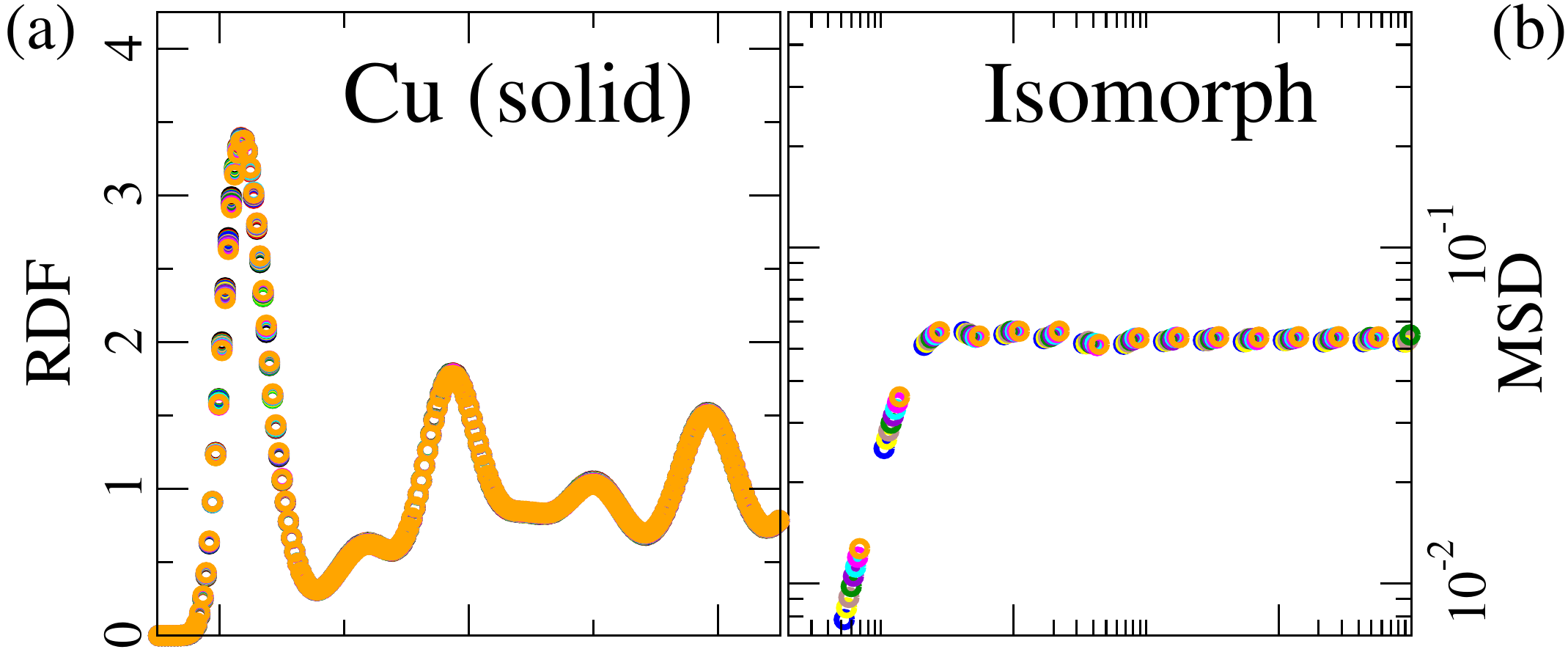}
   \\
   \includegraphics[width=\linewidth]{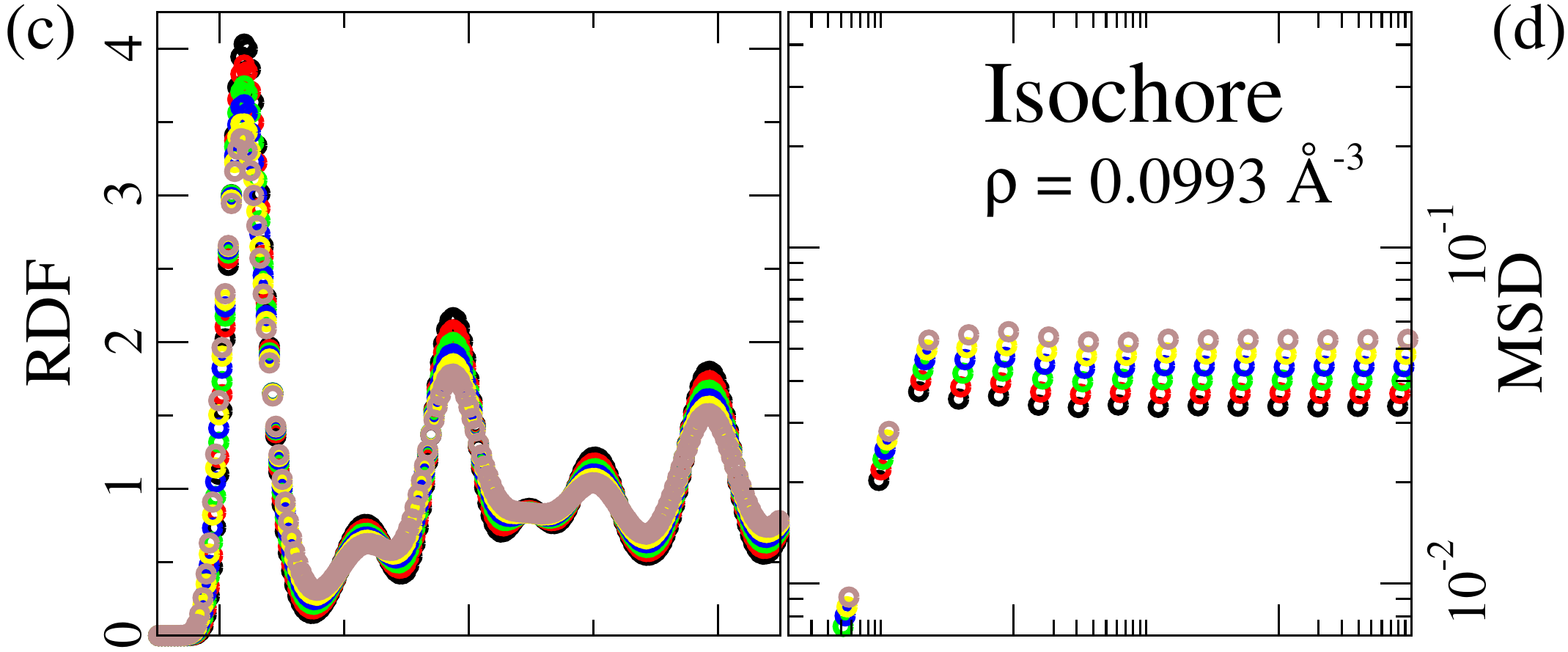}
   \\
   \includegraphics[width=\linewidth]{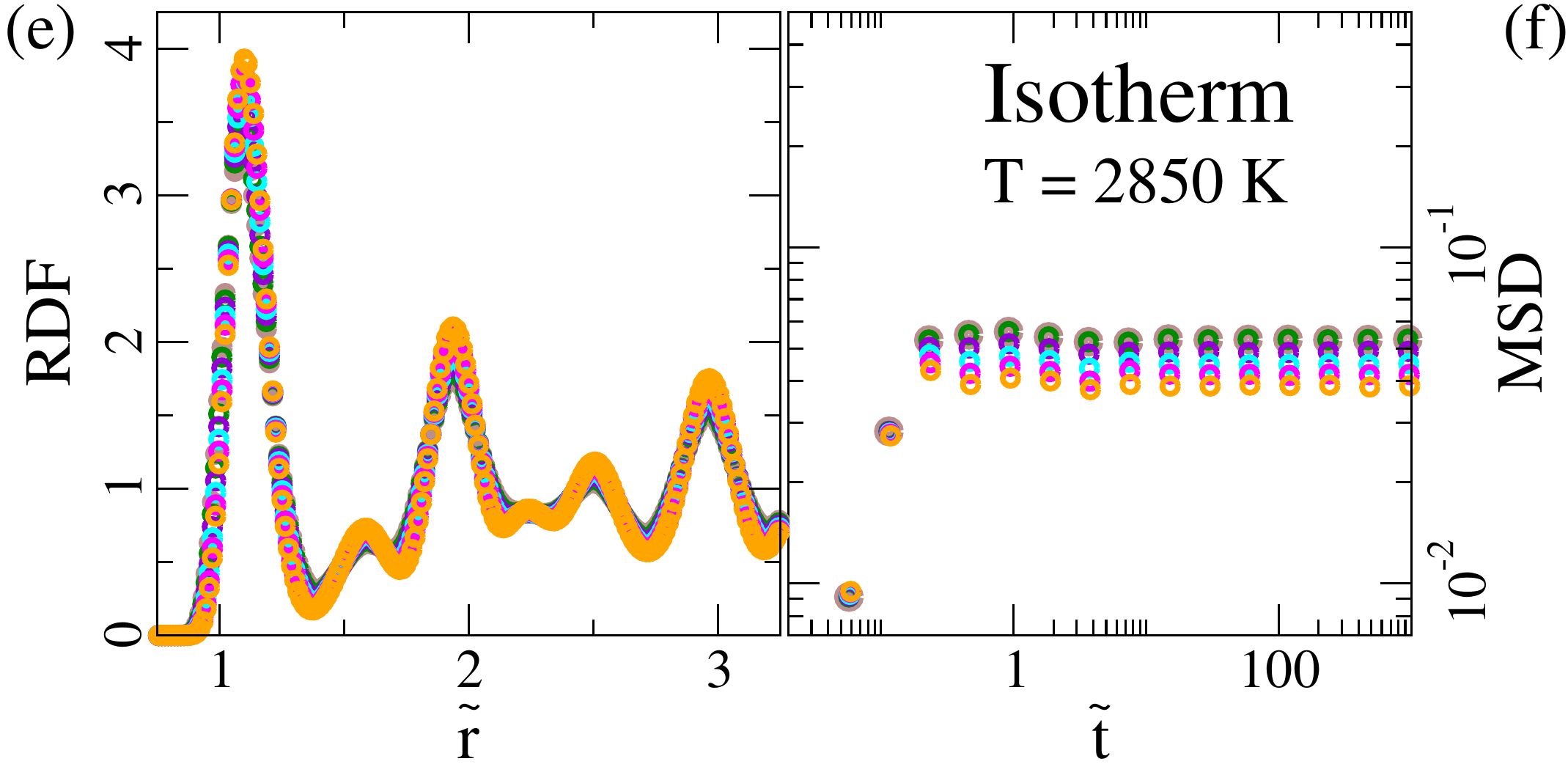}
   \\
   \caption{RDF and MSD along various iso-curves in the solid phase of EMT-Cu (Fig. 3). The RDF has been plotted against the reduced pair distance $\tilde r\equiv \rho^{1/3}r$, while the MSD has been put into reduced form by multiplying with $\rho^{2/3}$ and plotted against the reduced time $\tilde t\equiv t/(\rho^{-1/3}\sqrt{m/k_B T})$. Each sub-figure contains multiple curves, one for each state point on the iso-curve in question. The colors indicate the state point according to Fig.~\ref{fig:Cu_phase_diagram}. Both structure and dynamics are invariant to a good approximation along the isomorph.}
   \label{fig:Cu_rdf+msd_solid}
\end{figure}

\begin{figure}
\centering
   \includegraphics[width=\linewidth]{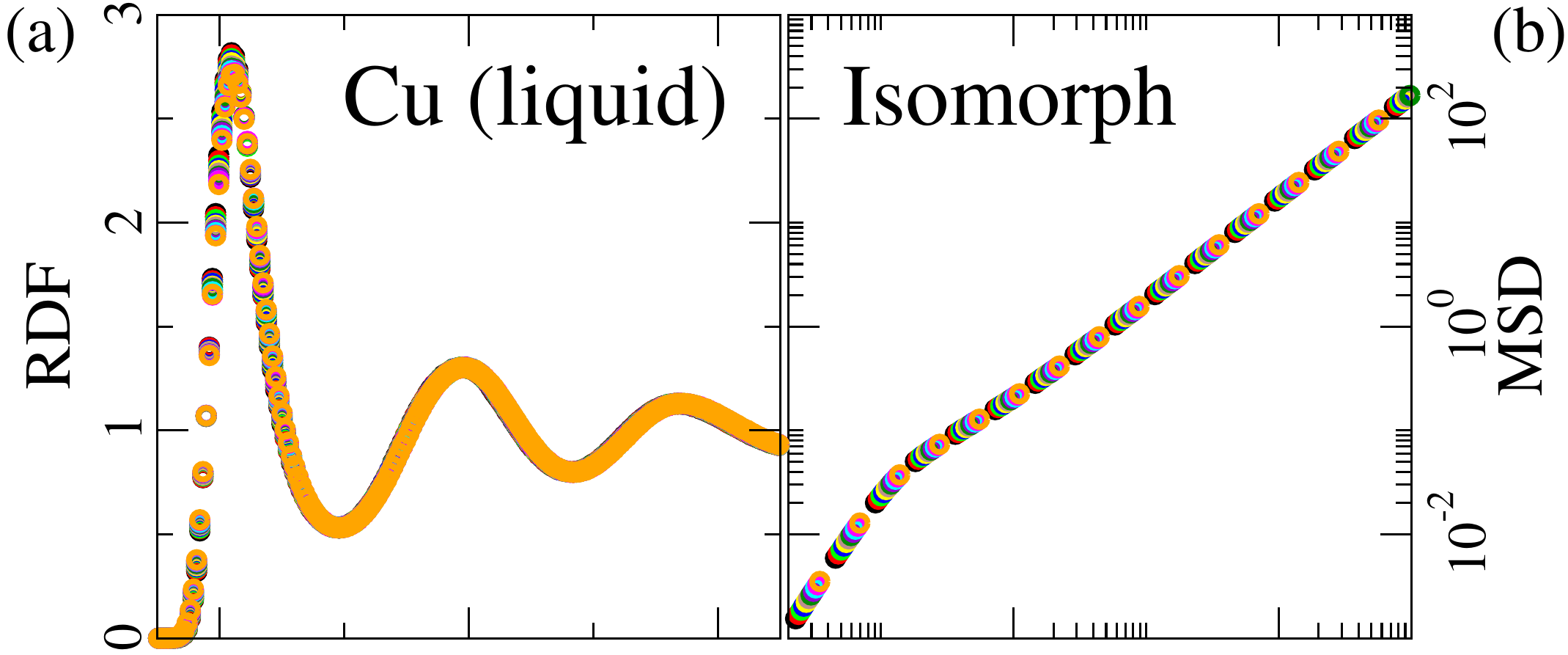}
   \\
   \includegraphics[width=\linewidth]{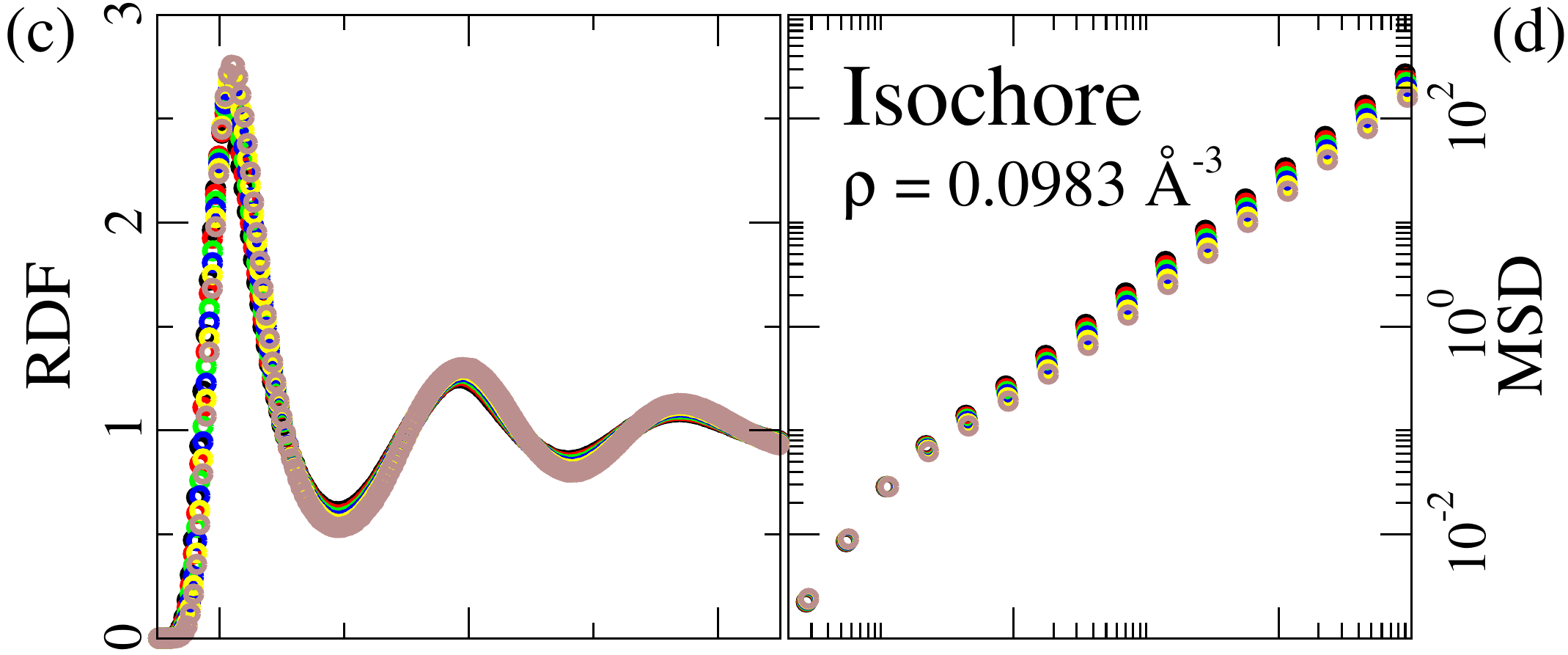}
   \\
   \includegraphics[width=\linewidth]{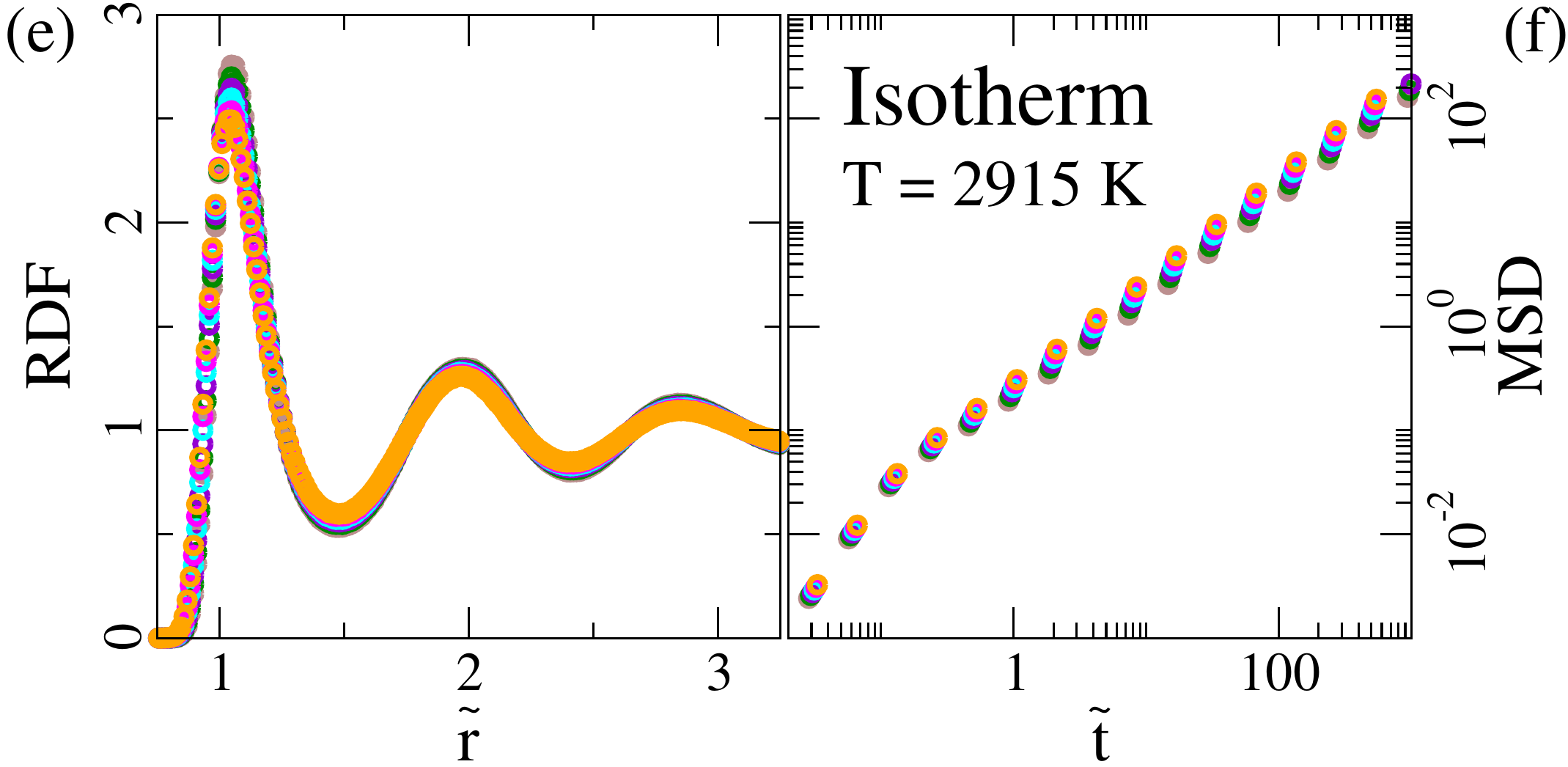}
   \\
   \caption{RDF and MSD along iso-curves in the liquid phase of EMT-Cu. Details are as in Fig.~\ref{fig:Cu_rdf+msd_solid}. Both structure and dynamics are invariant to a good approximation along the isomorph.}
   \label{fig:Cu_rdf+msd_liquid}
\end{figure}

\subsection{Isomorphs in EMT-Cu}

The starting point for these calculations is a known point on the melting curve to serve as the reference state point. This was determined using the interface pinning method\cite{Pedersen:2013} at 2008\,K. This method returns the pressure for a given temperature on the melting curve, along with the volumes (densities) of each phase. Single-phase $NVT$ simulations were then run at those densities and that temperature to generate the two isomorphs using the DIC. Note that the interface-pinning method requires simulations of a so-called $NP_zT$ ensemble, which maintains a fixed pressure in one direction (the z-direction) and fixed area in the remaining two. The $NVT$ simulations in RUMD were realized using the Nos\'e-Hoover thermostat\cite{Nose:1984,Toxvaerd:1991} while the interface pinning simulations employed Langevin dynamics.\cite{Allen/Tildesley:1987} For the solid phase a simulation box with 2048 particles was simulated, which corresponds to an FCC crystal with $8 \times 8 \times 8$ cubic unit cells. The same number of particles was used for the liquid phase. The isomorphs were determined using the DIC, increasing density by up to ~23\%. Figure~\ref{fig:Cu_phase_diagram} shows the mapped liquid and crystal isomorph for Cu, as well as the state points that were simulated on the isomorph of each phase. For comparison, isochores, and isotherms sharing, respectively, the density or the temperature of the middle state point are also simulated for each phase as indicated.

An indication of the degree of isomorphism can be obtained by using the scaled configurations to estimate the correlation coefficient $R$ and density-scaling exponent $\gamma$ at the points along the isomorph. These, along with the temperatures obtained from the DIC, are plotted versus the density in Fig.~\ref{fig:Cu_DIC_consistency}. The density-scaling exponent $\gamma$ decreases monotonically with increasing density and temperature along the isomorph. This behaviour is consistent with previous experience from the EMT metals \cite{Friedeheim/others:2019} and seems to be a general feature of metals.\cite{Hu/others:2016} Consequences of this density dependence of $\gamma$ for melting will be discussed below. A consistency check, first used in Ref.~\onlinecite{Friedeheim/others:2019},  can be done by fitting the $\gamma$ values to the function $\gamma(\rho) = 
a + b/\rho^n$. This function can be integrated analytically and gives an alternative estimate of the temperature along the isomorph. The middle panel of the figure shows that this estimate agrees well with the temperatures obtained from the DIC. In the bottom panel the correlation coefficients are plotted. They all exceed 0.97, indicating a strong degree of isomorphism.

To explicitly demonstrate the quality of the isomorphs we have run simulations at the state points determined by the DIC along the solid and liquid isomorphs. Figures~\ref{fig:Cu_rdf+msd_solid} and \ref{fig:Cu_rdf+msd_liquid} show structural and dynamical data for Cu in the solid and liquid phases, respectively. Similar data for the solid phase were presented in Ref.~\onlinecite{Friedeheim/others:2019}. Each figure shows radial distribution function (RDF) and mean-squared displacment (MSD) data expressed in reduced units for the isomorphs and, for comparison, also for the isotherms and isochores indicated in Fig.~\ref{fig:Cu_phase_diagram} by the corresponding colors. The plots for the state points on the isomorphs are well on top of each other, while they show noticeable variation for the state points along an isochore. The RDFs of state points along the solid isotherm are also very similar to each other, reflecting simply the validity of the harmonic approximation for crystals. Thermodynamic data for the solid and liquid Cu isomorphs are listed in Tables~\ref{tab:Cu_solid}, \ref{tab:Cu_liquid} in Appendix~\ref{app:thermodynamic_data}.

\begin{figure}
    \centering
    \includegraphics[width=0.8\linewidth]{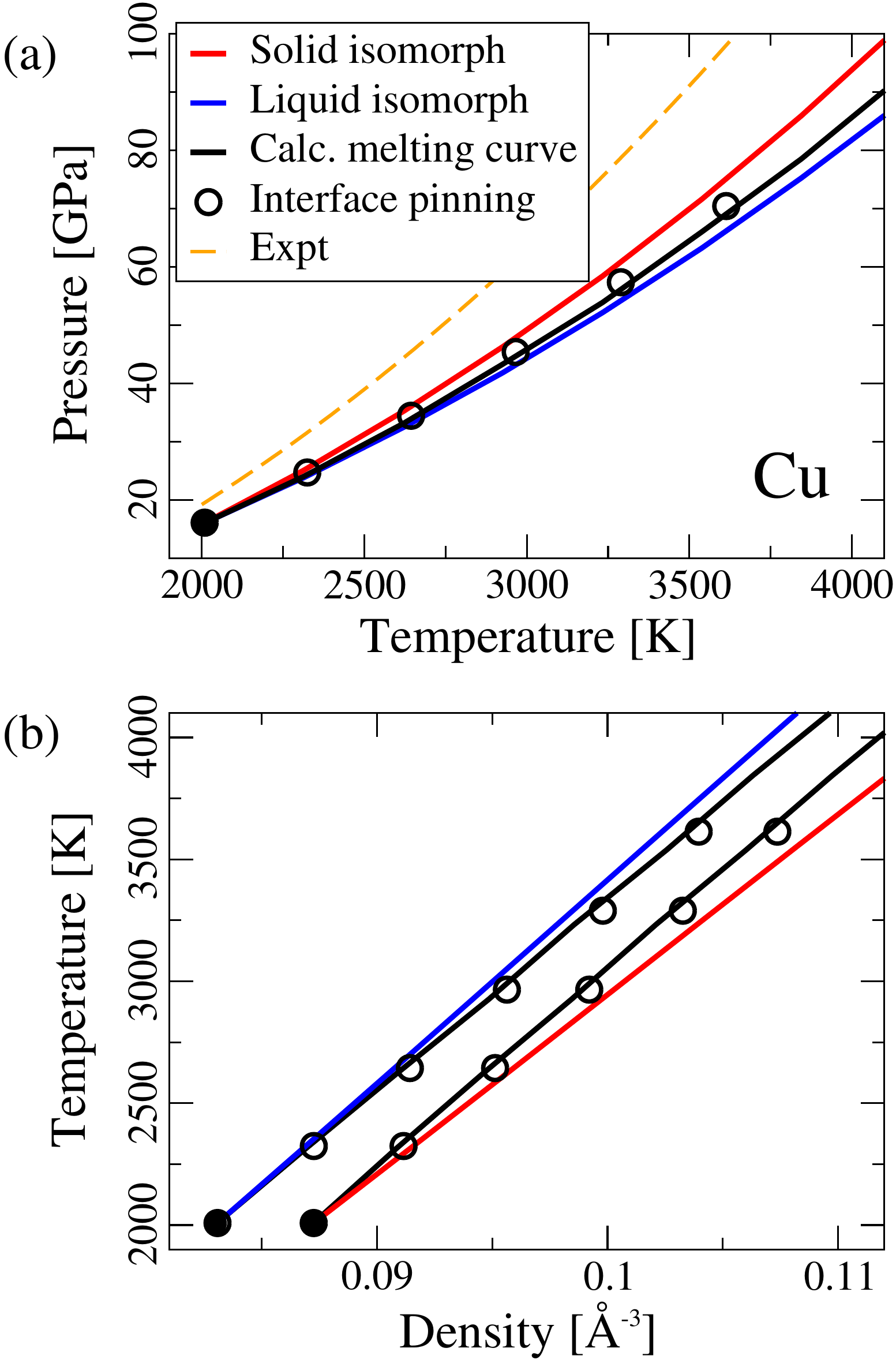}
    \caption{Comparison of the predicted (dashed) melting curves with coexistence state points determined using interface pinning (open circles) for EMT-Cu. The melting curve is predicted exclusively from simulations at the reference state point, one in the liquid and one in the solid phase. The reference state point is marked by the full circle. (a) shows the pressures along the solid and liquid isomorph generated from the reference point with the melting pressure. (b) shows the same in the $(\rho,T)$ diagram. For comparison the experimental melting of Errandonea\cite{Errandonea:2013} shown as the orange dashed line.}
    \label{fig:EMT_results}
\end{figure}

\subsection{Melting curve for EMT-Cu}

High pressure experimental data for the Cu melting curve has been reported by, for example, Japel {\em et al.},\cite{Japel/others:2005} and Errandonea.\cite{Errandonea:2010,Errandonea:2013} Corresponding computational studies have used both ab initio methods\cite{Vocadlo/others:2004} and empirical potentials\cite{Belonoshko/others:2000, Wu/others:2011, Ghosh:2012} There is broad agreement between the experimental and computational methods, which is consistent with the electronic structure being stable due to the filled d—bands.\cite{Japel/others:2005} Recent computational studies have, however, indicated that a bcc phase appears at sufficiently high pressures and temperatures\cite{Smirnov:2021,Baty/Burakovsky/Errandonea:2021}. Our calculated melting pressure and resulting freezing and melting densities are shown in Fig.~\ref{fig:EMT_results}. For comparison, coexistence state points for the same system determined using interface pinning are also included. The melting pressure predicted from the isomorph-based method agrees well with the interface points, although not as perfectly as in the Lennard--Jones case of Ref.~\onlinecite{Pedersen/others:2016a}, where analytic expressions for potential energy, virial, and temperature along isomorphs are available. A small systematic shift can be observed with the calculated melting pressure being slightly below the interface pinning results; the same applies also for the predicted densities of liquid freezing and solid melting. Nevertheless, Fig. \ref{fig:EMT_results} demonstrates that the method works well also for systems for which no analytic expressions are available for the isomorphs or for how thermodynamic quantities vary along them. The melting curve shows good qualitative agreement with the experimental data of Errandonea\cite{Errandonea:2013}. Quantitatively, the EMT curve is around 10\% higher in temperature, compared to the experimental data. This deviation largely stems from the EMT model of Cu, rather than from isomorph theory. Our point is that, given any computational model, such as EMT or DFT, isomorph theory can generate very accurate melting curves at considerably reduced computational cost.

All data required to produce the melting and freezing line predictions can be obtained from simulations at only the reference state point ($T=2008$\,K). Combining the isomorph-based prediction method, Eqs.~(\ref{eq:meltingP}) and (\ref{eq:meltingrho}), with the DIC thus makes it possible to predict the melting and freezing curves in systems where conventionally used methods are too computationally expensive. In the following section this is tested for DFT simulations of Aluminum.

\section{Isomorphs and melting curve for DFT Aluminum}\label{sec:isomorph_invariance_DFT}

We now apply the melting curve method expressed through Eqs.~(\ref{eq:meltingP}) and (\ref{eq:meltingrho}) to data from density functional theory simulations carried out using the Vienna {\em ab-initio} simulation package (VASP) \cite{Kresse/Hafner:1994,Kresse/others:1996a,Kresse/others:1996b}.

\subsection{\label{sec:DFT-isomorphs}DFT simulations of isomorphs}

First principles or \textit{ab initio} methods try to retain materials properties directly from the Sch\"odinger equation, involving only the physical constants as experimental input. This is very hard in practice. The computational complexity of conventional numerical solution techniques scales exponentially with system size, making calculations of more than a handful of electrons intractable even on the most powerful computing facilities.
{\bf
State\-/of\-/the\-/art \emph{ab initio} calculations employ a number of approximations that hold for
a wide range of temperatures and densities:
(i) the Born--Oppenheimer approximation, which separates the nuclear from the electronic quantum degrees
of freedom due to the nuclei being much heavier than the electrons. Typical time scales of the nucleus and
the electron dynamics are well separated and the interactions between nuclei and electrons are well described
by mean\-/field interactions. Expectation values of observables of one system enter only as parameters
for the Hamiltonian of the respective other system.
(ii) the adiabatic approximation, assuming that the relaxation of the electrons into their thermal
equilibrium is so fast that only the positions of the nuclei enter the electronic Hamiltonian as
parameters but not their velocity or their acceleration.
(iii) the density of the nuclei is low compared to their de Broglie wavelength and one can treat the
dynamics of the nuclei classically.

Under these assumptions, the nuclear problem reduces to the classical dynamics of point masses in an effective
potential energy landscape $U(\vec R)$ that is given by the free energy of the electrons
in the setting, where the nuclei are fixed to at the positions $\vec R$.
The electronic free energy
is a functional of the electronic equilibrium density only, a much simpler object than the entire electronic equilibrium density matrix.%
\cite{Mermin65}
Modern density functional theory
implementations employ the Kohn--Sham framework to retrieve accurate kinetic energy contributions
and rely on density and density-gradient dependent fits of the remaining so-called exchange\-/correlation contribution,
retrieved from high accuracy calculations of only a small number of electrons.%
\cite{Kohn/Sham:1965}
We approximate the temperature\-/dependent free energy functional by applying Fermi smearing
to Kohn--Sham orbitals found from a zero\-/temperature ground\-/state density functional.\cite{Kresse/Hafner:1994}
To this end, we set the Fermi smearing
energy to $k_\mathrm{B}T$. Note that
we do not use smearing to accelerate the
convergence of the orbital solver but rather
to approximate the free energy of electrons
at non\-/zero temperature.
}

The results of the present work have been obtained using the standard approximation by Perdew--Burke--Ernzerhof (PBE) of the exchange-correlation functional \cite{Perdew/others:1996} implemented by VASP using the projector augmented wave (PAW) method.\cite{Blochel:1994} Aluminum has only three valence electrons, which enables an efficient simulation using a frozen-core type pseudopotential. This means that only the outer shell electrons are treated explicitly while the inner electrons are considered frozen.
To verify the validity of this approximation under the extraordinarily high densities considered in this work, energy and virial of several hundred uncorrelated configurations of the MD trajectories have also been computed with eleven valence electrons per atom, freezing only the innermost two electrons. No statistically relevant difference was found between calculations with three and eleven valence electrons per atom.
The Brillouin zone was sampled at the $\Gamma$ point $(0, 0, 0)$. The initial choice for the kinetic energy cutoff for the plane-waves was 220\,eV. This value applies to the reference state point; in order to avoid discontinuities, the number of plane waves is kept constant when scaling configurations. This means that for the density changes of the DIC, the energy cutoff is scaled as 

\begin{align}
    E_{\mathrm{cut}} (\rho) =E_{\mathrm{cut}} (\rho_i)  \left(\frac{\rho}{\rho_i}\right)^{2/3}
\end{align}
where $\rho_i$ and $E_{\mathrm{cut}} (\rho_i)$ are, respectively, the initial density and the energy cutoff chosen for that density. The $\rho^{2/3}$ dependence is due to the plane-wave energy being proportional to wave number squared, while the wave number of any given plane wave scales inversely with the linear size of the box, and thus as $\rho^{1/3}$. The simulation box for Al contained 108 atoms for each phase. This number is determined by the crystal phase having an FCC structure with 108 atoms for a crystal size of $3\times3\times3$ cubic unit cells. The $NVT$ molecular dynamics (MD) simulations were carried out using a Langevin thermostat.

Each simulation run consists of 50000 MD steps with a time step of 2\,fs. The first 20000 steps ensure that the system reaches equilibrium in the respective phase. From the remaining 30000 steps every 100th configuration is sampled for the DIC scaling procedure. The velocity auto-correlation functions indicate sufficient statistical independence after 100 MD steps. For determining the isomorphs, new densities were obtained from scaling the initial liquid and solid configurations increasing their density by the following factors: $1.1, 1.2, \ldots, 1.9, 2.0$.
To obtain data on structure and dynamics for testing isomorph invariance, we have selected the relative densities $1.3, 1.6$, and $1.9$ and simulated at the temperature predicted from the DIC for the given density. These simulations also consisted of 50000 steps with a time step of 2\,fs. Configurations from the last 20000 steps of the simulation were used to calculate the radial distribution and the velocity auto-correlation functions. In addition to the scaling of configurations necessary for the DIC, we also carried out much smaller scaling around the specified densities to determine the hypervirial as a numerical derivative (the hypervirial is necessary to determine the isothermal bulk modulus in \eqref{eq:K_T_formula}).

\subsection{Reference state point for Al isomorphs}

As for Cu, a state point at solid-liquid coexistence is chosen as reference. We use literature data for the melting curve from experiments and other DFT simulations, both to obtain a reference point and to validate our predictions. The reference point for Al was the experimentally determined coexistence point at temperature 2265\,K and pressure 27.5\,GPa.\cite{Boehler/Ross:1997,Hanstrom/Lazor:2000}
Interface pinning DFT calculations for Al are beyond the scope of this work and have not been conducted. DFT calculations scale with system size as $\mathcal O(N^3)$. Thus the two-phase calculation in the interface pinning method with 216 atoms is more than 8 times slower than the single-phase calculations with 108 atoms used in the DIC. Furthermore, non-cubic two-phase calculations exhibit strong pressure anisotropies due to different length scales of the periodic boundary conditions, to which the conduction electrons of metals are sensitive. \cite{Pedersen/others:2013,Friedeheim:2020}. The densities of the two phases at this reference state point were not reported in Refs. \onlinecite{Boehler/Ross:1997,Hanstrom/Lazor:2000}; these are, however, needed for the DIC calculations in the $NVT$ ensemble. The corresponding densities were estimated by simulating a series of different box sizes to find the box size at which the pressure matched the desired pressure. The resulting liquid density was $0.0688\,\text{\AA}^{-3}$ and the crystal density was $0.0714\,\text{\AA}^{-3}$, see also Table~\ref{tab:Al_data}. At these densities the simulated average pressure matched the desired pressure to within 0.5-0.6\,GPa.


From the reference state point a solid and a liquid isomorph were generated. For comparison, points along an isotherm and an isochore in each phase were studied as well. The part of the phase diagram indicating these state points is shown in Fig.~\ref{fig:Al_phasediagram}. The lines represent the isomorphs found from applying the DIC to data from a simulation at the reference points marked by the solid dots. The other points noted in the figure correspond to state points where additional simulations were carried out. Including the reference points, this means that for each phase a total of four isomorphic state points and three state points that are isothermal or isochoric to state points along the isomorph were simulated. In the figure, isomorphic state points are referred to by their label according to the change in density relative to the reference point. Even though the initial temperatures are the same for the liquid and solid phases and the corresponding densities were scaled by the same factors, the obtained temperatures are different, with the solid temperature being lower than the liquid temperature for the same density-scaling factor.
This is a consequence of the scaling exponent $\gamma$ decreasing with increasing density, as seen in Fig.~\ref{fig:Al_DIC_check}(a), and the solid having a higher initial density.

\subsection{Isomorph invariance in Al}

\begin{figure}
    \centering
    \includegraphics[width=.75\linewidth]{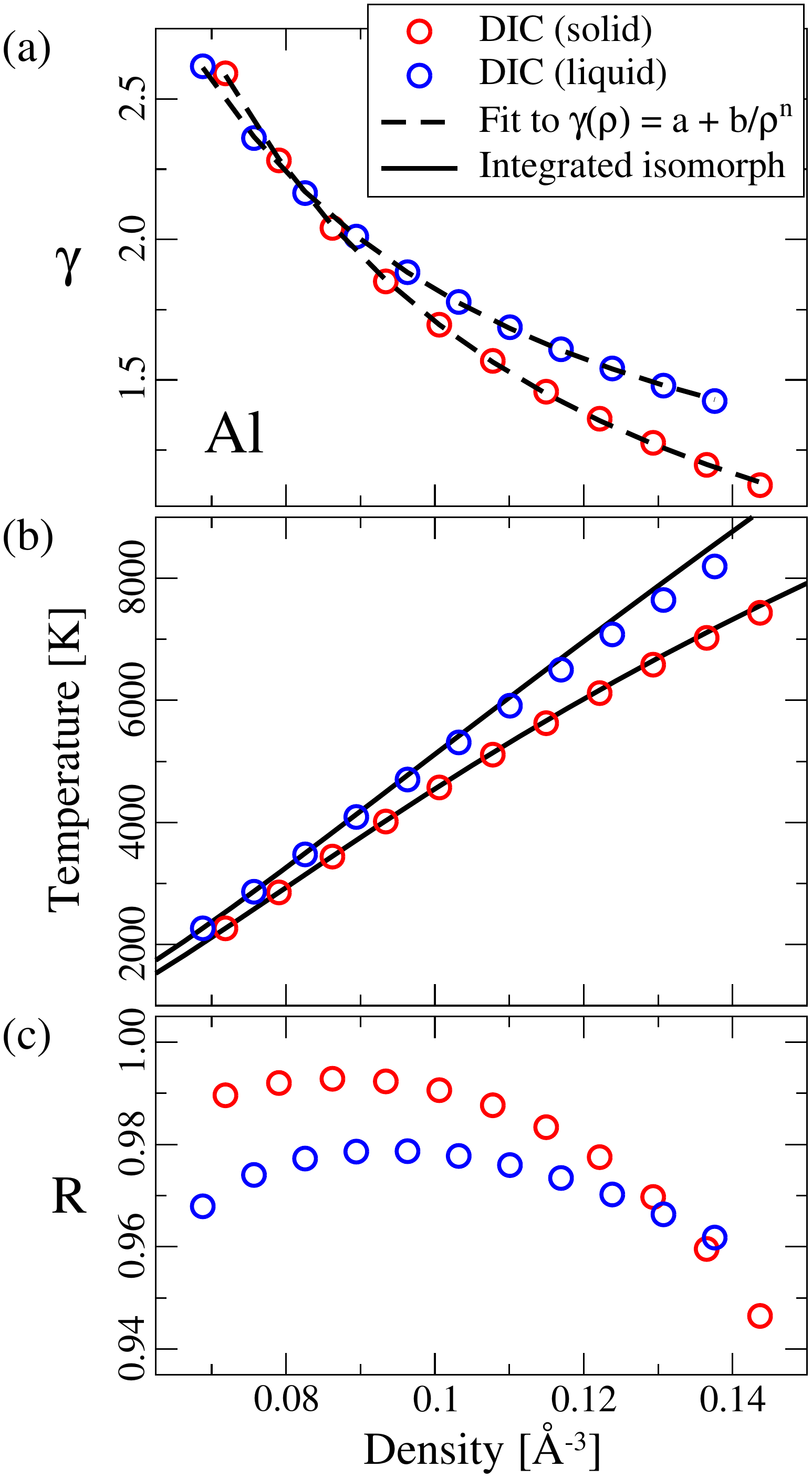}
    \caption{DIC self consistency check for Al. Open circles represent data for the isomorphs from DIC, red for the solid and blue for the liquid. The dashed line in the top panel was obtained from fitting to the DIC points. The parameter for the fitted expression for $\gamma(\rho)$ are: $a=0.3422, b=0.04338, n=1.498$ for the solid and $a=0.8181, b=0.02826, n=1.551$ for the liquid isomorph. The resulting integrated isomorph is given by the solid line in the middle panel. The bottom panel shows the correlation coefficient for both isomorphs obtained from the DIC.}
    \label{fig:Al_DIC_check}
\end{figure}

 Figure~\ref{fig:Al_DIC_check} shows the scaling exponent $\gamma$, the actual DIC-generated isomorphs in the density-temperature diagram, and the virial potential-energy correlation coefficient $R$. We emphasize that $R$ and $\gamma$ were not obtained by simulating at the isomorphic state points, but purely by scaling sampled configurations from the reference simulation, computing their energies and virials, and then applying Eqs.~(\ref{eq:R}) and ~(\ref{eq:gamma}), respectively. As in the case of EMT-Cu, $\gamma$ decreases monotonically with increasing pressure/temperature along the isomorph. For all state points along the isomorph the correlation coefficient is well above the $R>0.9$ threshold defining an R-simple system. This means that good isomorphs in this part of the phase diagram can be expected. We find the same behaviour as previously seen for other metallic systems\cite{Friedeheim/others:2019,Friedeheim/Dyre/Bailey:2021} of $R$ decreasing slightly as density and temperature increase. Given that $\gamma$ decreases substantially and presumably reaches zero,
$R$ must also eventually decrease since the correlation necessarily goes to zero when $\gamma$ does, as seen from Eqs.~(\ref{eq:R}) and (\ref{eq:gamma}). The Cu data do not show this decrease in Fig.~\ref{fig:Cu_DIC_consistency}, but the density range there is smaller; a decrease may be expected to occur for Cu at higher densities. $R$ begins also to decrease at low density, where the configurational part of the pressure approaches zero, and strong correlation typically breaks down.\cite{Bailey/others:2013}

\begin{figure}[t!]
    \centering
    \includegraphics[width=0.8\linewidth]{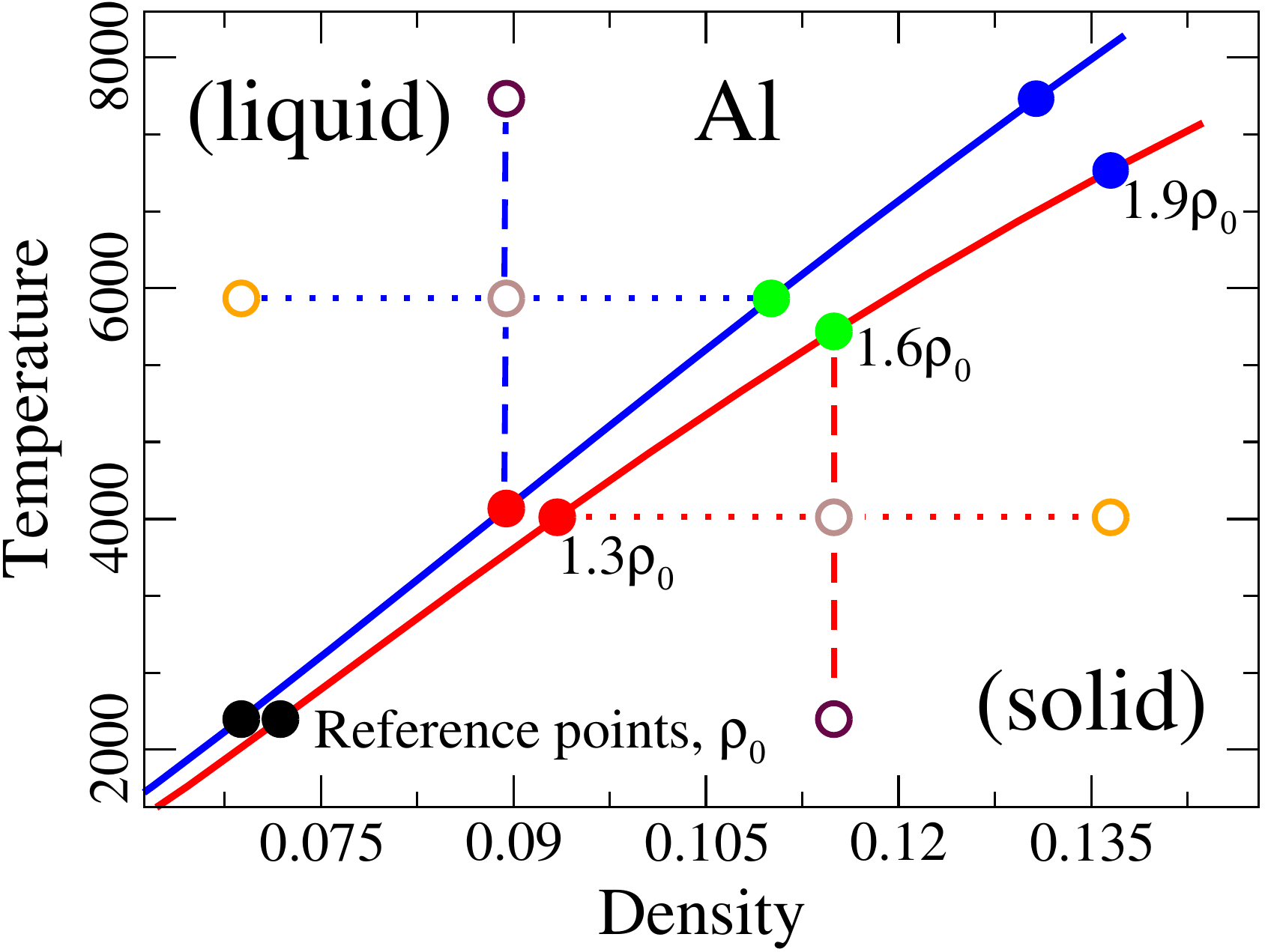}
    \caption{Relevant part of the Al phase diagram marking all points studied by DFT. The solid lines denote isomorphs generated from the reference points (black circles). Additional simulations were carried out at the state points marked by filled circles. The points connected by the dotted/dashed line are on the same isotherm/isochore, respectively. Iso-curves on the liquid side of the phase diagram are given in blue and in red for the solid side. State points along the isomorphs are labeled according to the relative change with respect to the reference point density. The colors of the filled circles correspond to the colors used in Figs.~\ref{fig:Al_rdf+vacf_solid} and \ref{fig:Al_rdf+vacf_liquid} for structure and dynamics data at the respective state points.
    }
    \label{fig:Al_phasediagram}
\end{figure}

\begin{figure}[b!]
\centering
   \includegraphics[width=\linewidth]{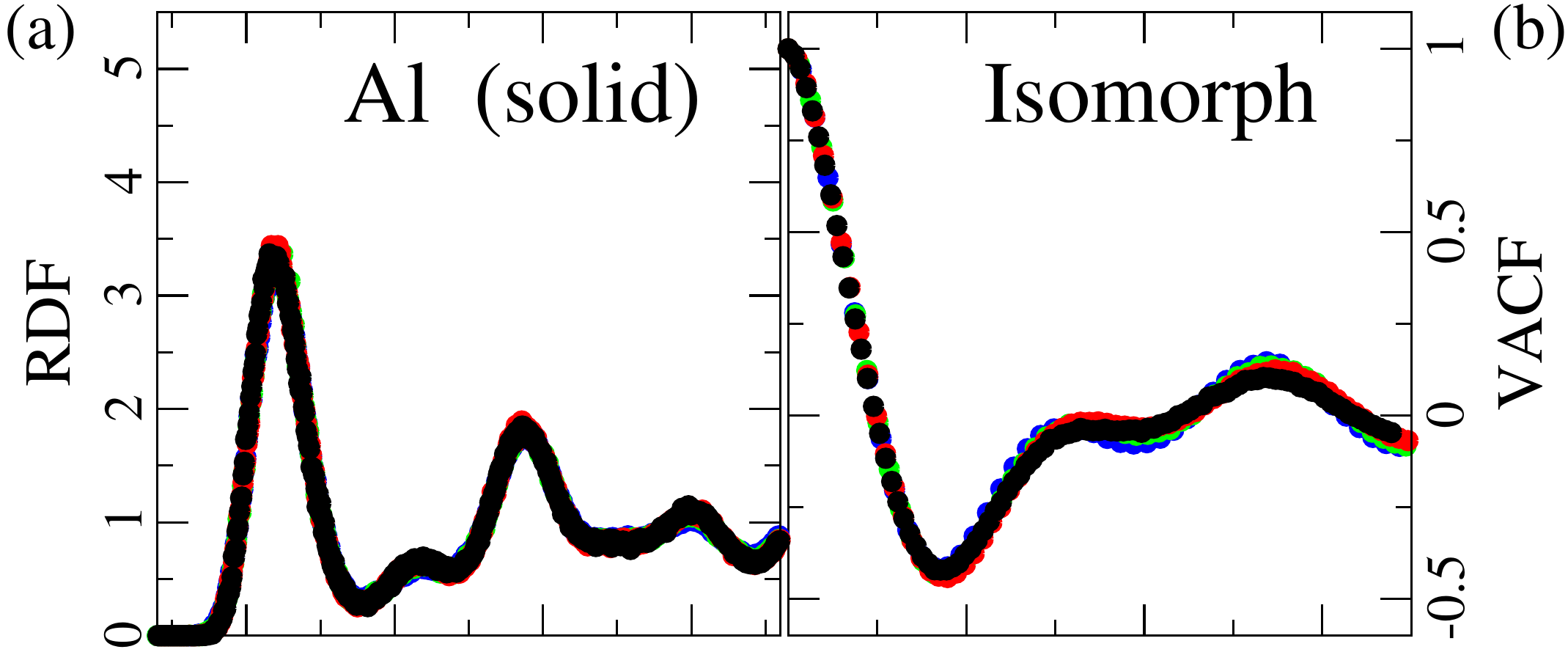}
   \\
   \includegraphics[width=\linewidth]{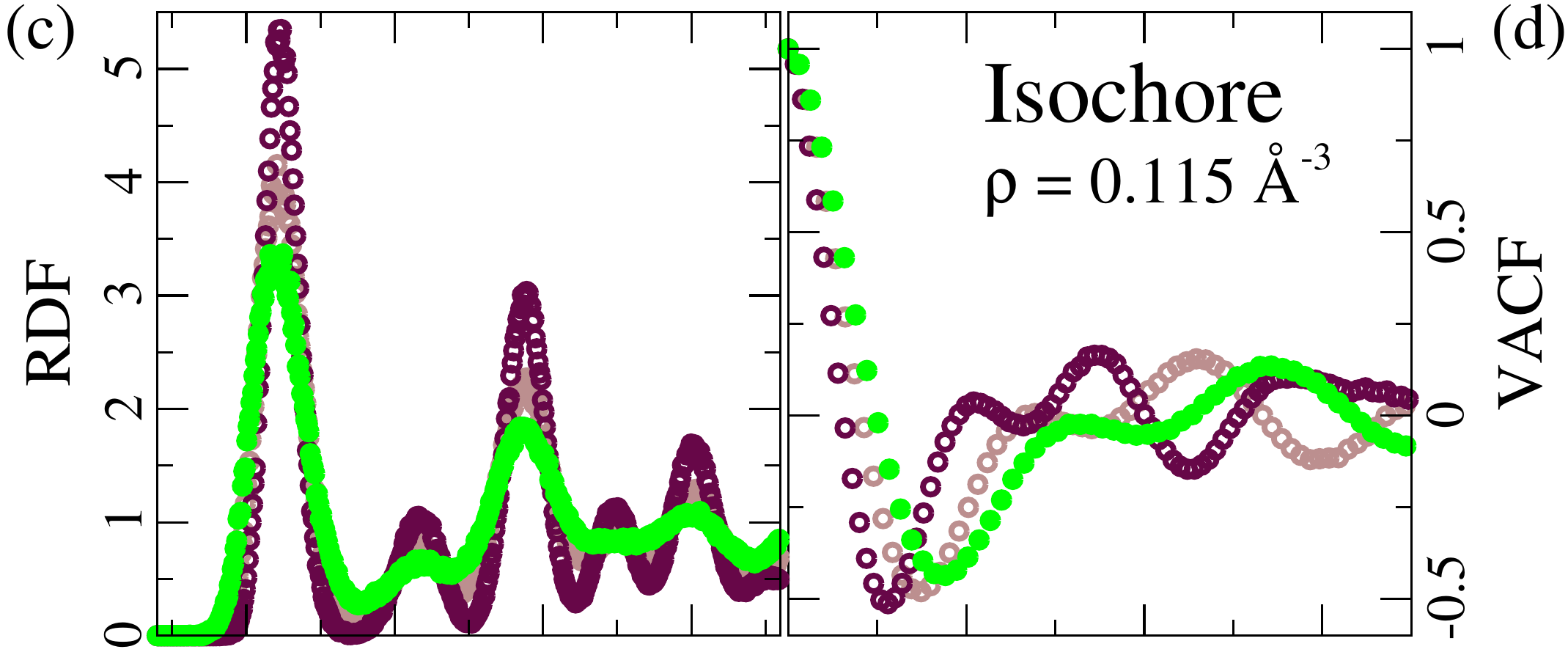}
   \\
   \includegraphics[width=\linewidth]{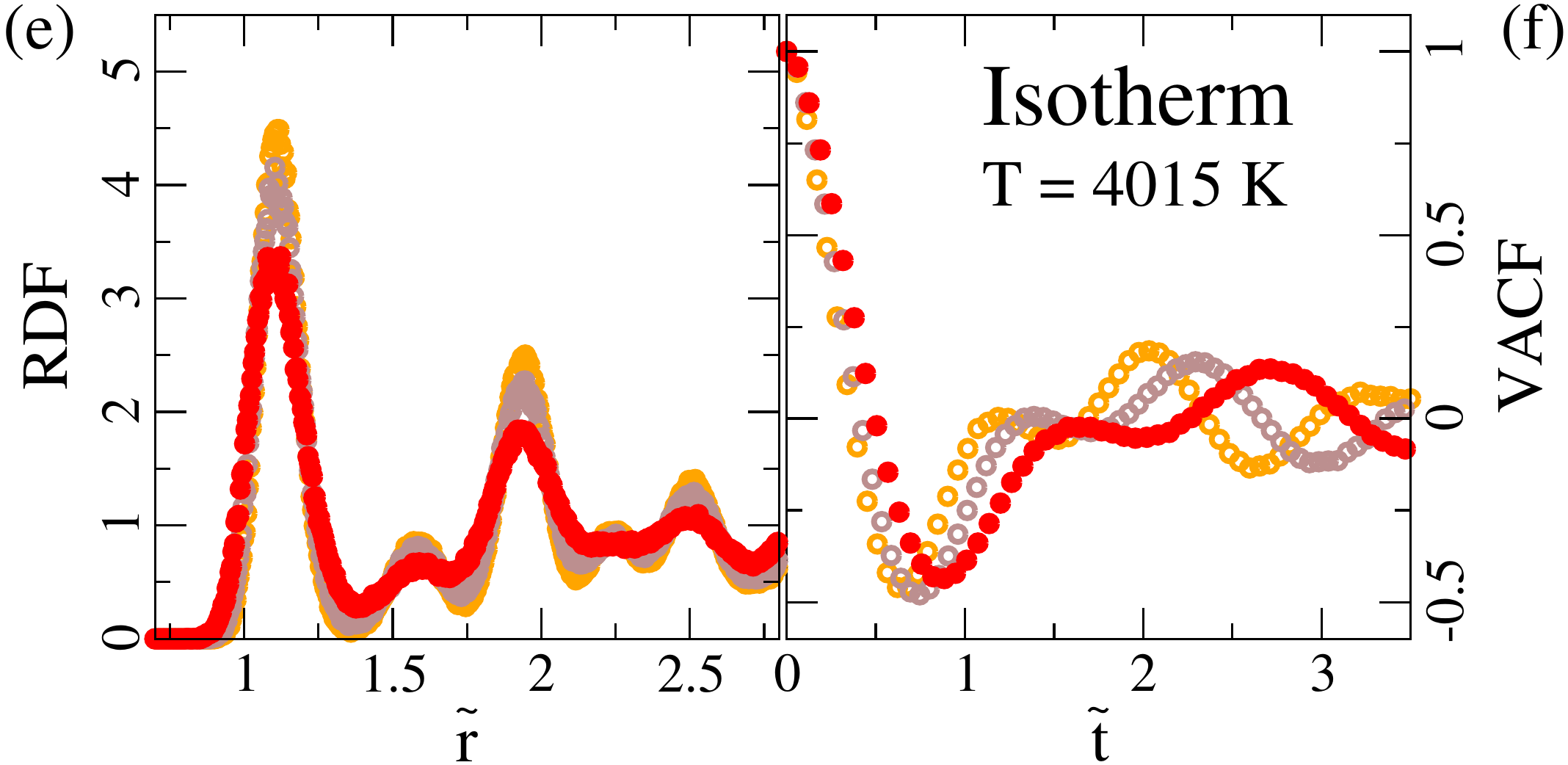}
   \\
   \caption{RDF and VACF along iso-lines in the solid side of the phase diagram of Al. Each panel shows the RDF on the left and VACF on the right for the same state points (values are given in Table~\ref{tab:Al_data}). The state points for the top row are along the same isomorph. State points shown in the middle and bottom panel are along an isochore and isotherm, respectively.}
   \label{fig:Al_rdf+vacf_solid}
\end{figure}

\begin{figure}[b!]
\centering
   \includegraphics[width=\linewidth]{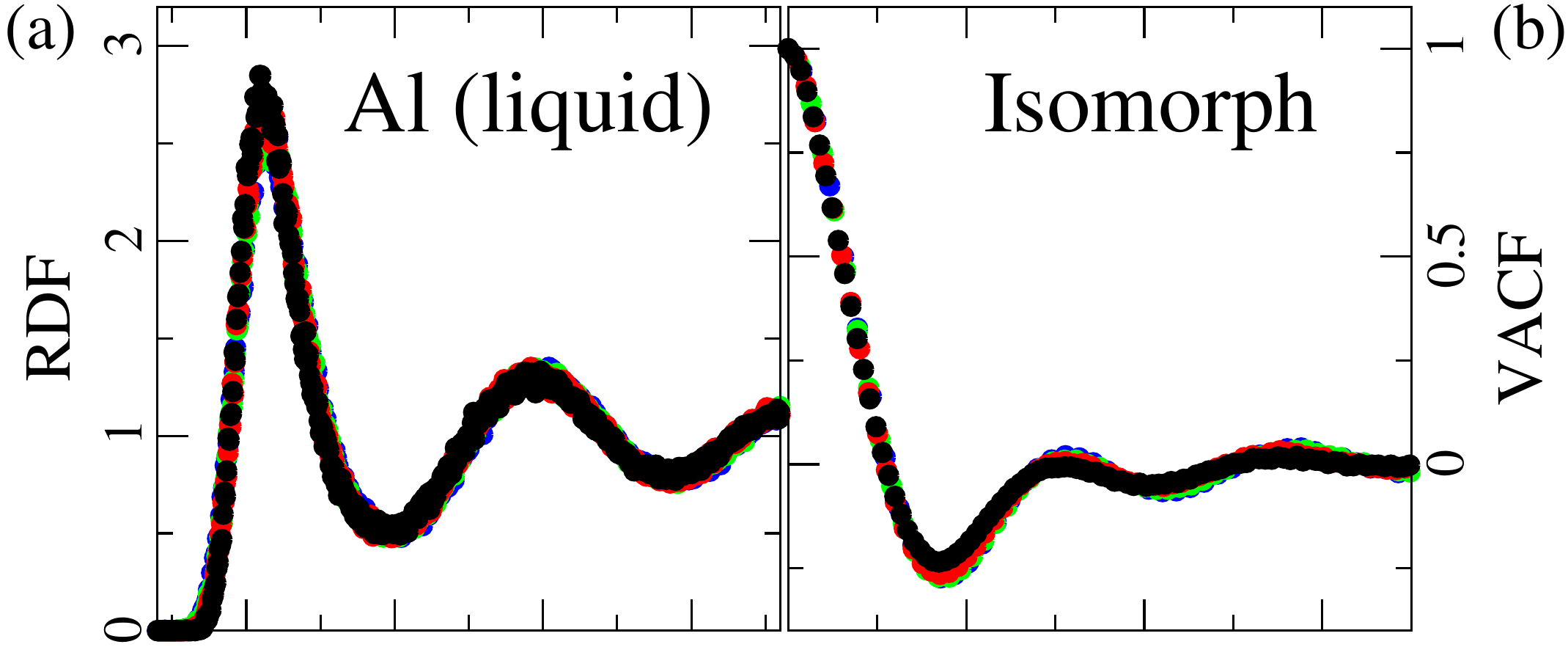}
   \\
   \includegraphics[width=\linewidth]{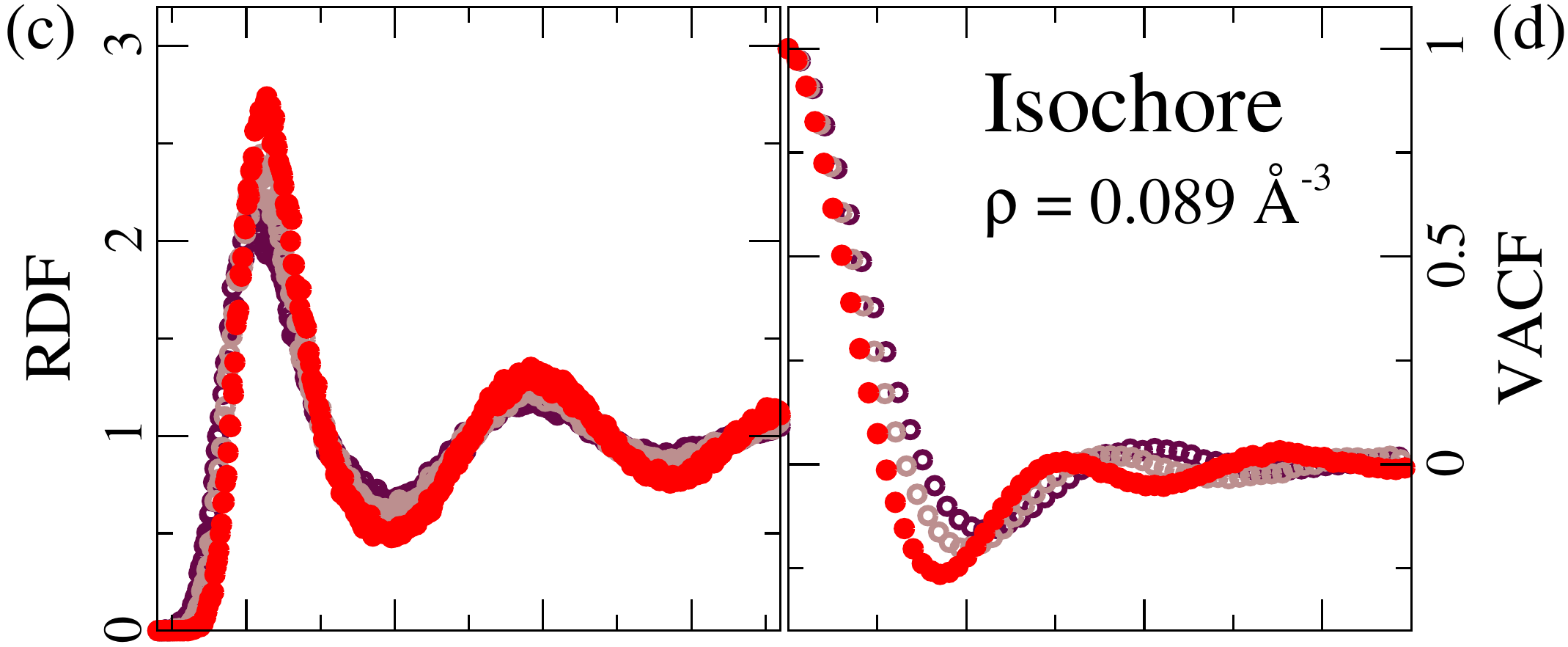}
   \\
   \includegraphics[width=\linewidth]{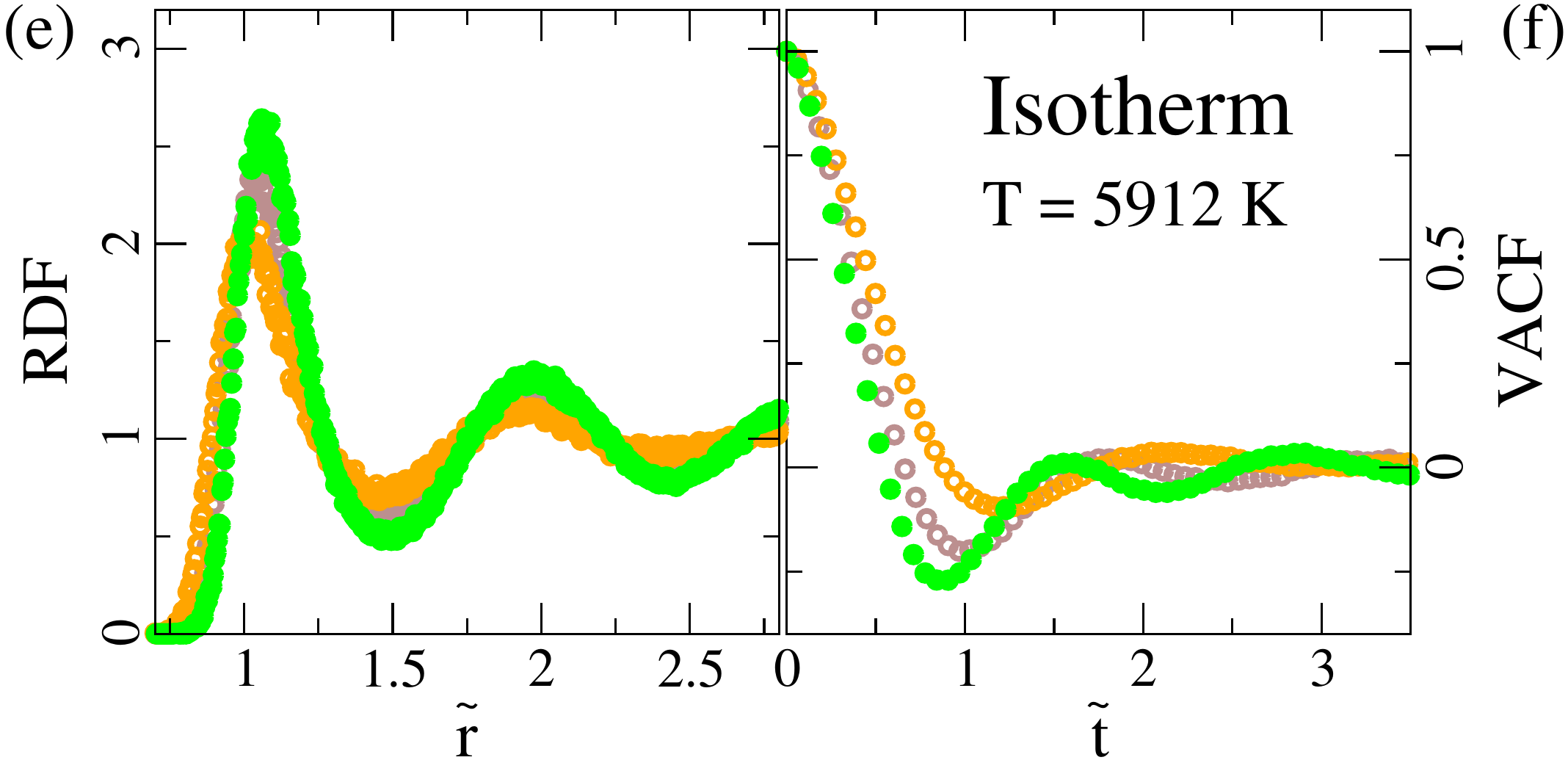}
   \\
   \caption{RDF and VACF along iso-lines in the liquid side of the Al phase diagram, similar to the presentation of the solid side in Figure~\ref{fig:Al_rdf+vacf_solid}. Each panel shows the RDF on the left and VACF on the right for the same state points.}
   \label{fig:Al_rdf+vacf_liquid}
\end{figure}

We show in Fig.~\ref{fig:Al_DIC_check} the same consistency check used for Cu, now applied to Al. Since the DIC is not exact and since the reference point for the DIC and the fitting is the lowest density-temperature state point, small deviations between the fitted and the DIC state points can be expected at the high density and high temperature end. The deviation between the mapped and the integrated isomorph is indeed small, again suggesting that good isomorph invariance can be expected. The overall similarity in the behavior of $\gamma$ and $R$ between DFT and EMT provides further evidence that the EMT potential, although simple compared to other many-body potentials, gives a good overall description of metallic interactions.

\begin{table*}
    \begin{tabular}{l l || c c c || c c | c c }
         $\rho/\rho_0$ & Color & $T$ [K] & $P$ [GPa] & $\rho$ [\AA$^{-3}$] & $\gamma_{DIC}$ & $R_{DIC}$ & $\gamma_{Sim.}$ & $R_{Sim.}$ \\[1mm] \hline \hline
         1.0 ($l$) & black & 2265 & 27.82 & 0.0688 & & & 2.616 & 0.9679 \\
         1.3       & red & 4090   & 94.92 & 0.0894 & 2.010 & 0.9786 & 1.935 & 0.9846 \\
         1.6       & green & 5912 & 198.7 & 0.1101 & 1.687 & 0.9760 & 1.558 & 0.9747 \\
         1.9       & blue & 7644  & 340.5 & 0.1307 & 1.480 & 0.9663 & 1.338 & 0.9653 \\[1mm]
         \hline \hline
         1.0 ($s$) & black & 2265 & 27.67 & 0.0714 & & & 2.591 & 0.9896 \\
         1.3       & red   & 4015 & 97.96 & 0.0934 & 1.851 & 0.9923 & 1.843 & 0.9916 \\
         1.6       & green & 5626 & 210.0 & 0.1150 & 1.458 & 0.9834 & 1.464 & 0.9852 \\
         1.9       & blue  & 7022 & 365.3 & 0.1365 & 1.198 & 0.9596 & 1.169 & 0.9526
    \end{tabular}
    \caption{Temperature, pressure, and density of the isomorphic state points simulated for Al with the color names corresponding to the colors in Figs.~ \ref{fig:Al_rdf+vacf_solid} and \ref{fig:Al_rdf+vacf_liquid}. The top half of the table is for the liquid isomorph, the bottom is for the solid isomorph. The last four columns give the $\gamma$ and $R$ values as predicted from the DIC versus the values measured from simulation.}
    \label{tab:Al_data}
\end{table*}

A direct way to check for isomorph invariance is to look at structure and dynamics studied by means of RDF and the velocity auto-correlation function (VACF), respectively (we chose the VACF rather than the MSD used above for Cu due to the more limited time range accessible in {\em ab initio} simulations). Additional simulations at the state points from the DIC were only carried out at the selected state points noted in Fig.~\ref{fig:Al_phasediagram}. The resulting RDFs and VACFs along various iso-curves are shown in Fig.~\ref{fig:Al_rdf+vacf_solid} for the solid side of the phase diagram and in Fig.~\ref{fig:Al_rdf+vacf_liquid} for the liquid side.

Panels (a) and (b) of both figures show the structure and dynamics along the respective isomorphs. For comparison, structure and dynamics along an isochore and an isotherm are included in the middle (c, d)and bottom (e, f) panels, respectively. The curves along the isomorph shown in the top panels are almost identical. This invariance is not perfect; see for example the second minimum in Fig.~\ref{fig:Al_rdf+vacf_solid}(b) and the first minimum of the VACF in Fig.~\ref{fig:Al_rdf+vacf_liquid} (b), but there is a clear contrast between the isomorphs on one hand and the isochores and isotherms on the other. Values for pressure, temperature, and density along the isomorphs are given Table~\ref{tab:Al_data}. Also given in the table are $\gamma$ and $R$ values determined by actually performing MD simulations at the state points in question, which can be compared to the DIC values determined by scaling configurations sampled at the reference state-point simulations. They show good agreement with each other.
From that and from the agreement of the RDFs and VACFs we conclude the presence of isomorphic state points.

\subsection{Al melting curve}\label{sec:melting_curves_DFT}

The melting curve of Al was measured by Jayaraman in 1963 up to 3 GPa,\cite{Jayaraman/others:1963} and in the range 12-80 GPa by Boehler and Ross,\cite{Boehler/Ross:1997} and H\"anstr\"om and Lazor.\cite{Hanstrom/Lazor:2000} More recent data from Errandonea has filled the gap up to 12 GPa.\cite{Errandonea:2010} Computational studies have involved both {\em ab initio} methods,\cite{Bouchet/others:2009,Robert/others:2015} and empirical potentials.\cite{Ghosh:2012} Recent DFT calculations by Hong and van de Walle\cite{Hong/Walle:2019} have extended the melting curve to unusually large pressures and temperatures, and confirmed the existence of a maximum melting temperature around 3500 GPa and 20000K.

The limited simulation cell size accessible to DFT leads to non-negligible fluctuations of thermodynamic quantities, in particular for the pressure. The predicted melting pressure according to \eqref{eq:meltingP} is very sensitive to deviations in the virial at the reference temperature. A significant offset at the starting point is acquired if the average pressures of the liquid and the solid phases do not exactly match the coexistence pressure. Even though the $NVT$ simulation cells for the solid and the liquid at the reference point are chosen to be at the same pressure, deviation between set and measured pressure of about 0.5\,GPa are to be expected in practice.
This deviation is amplified by \eqref{eq:meltingP}, resulting in an offset of approximately 16.5\,GPa. To compensate for this we have corrected the virial such that the pressures at the reference point both equal the desired pressure (appendix~\ref{app:pressure_correction}).

\begin{figure}
    \centering
    \includegraphics[width=0.8\linewidth]{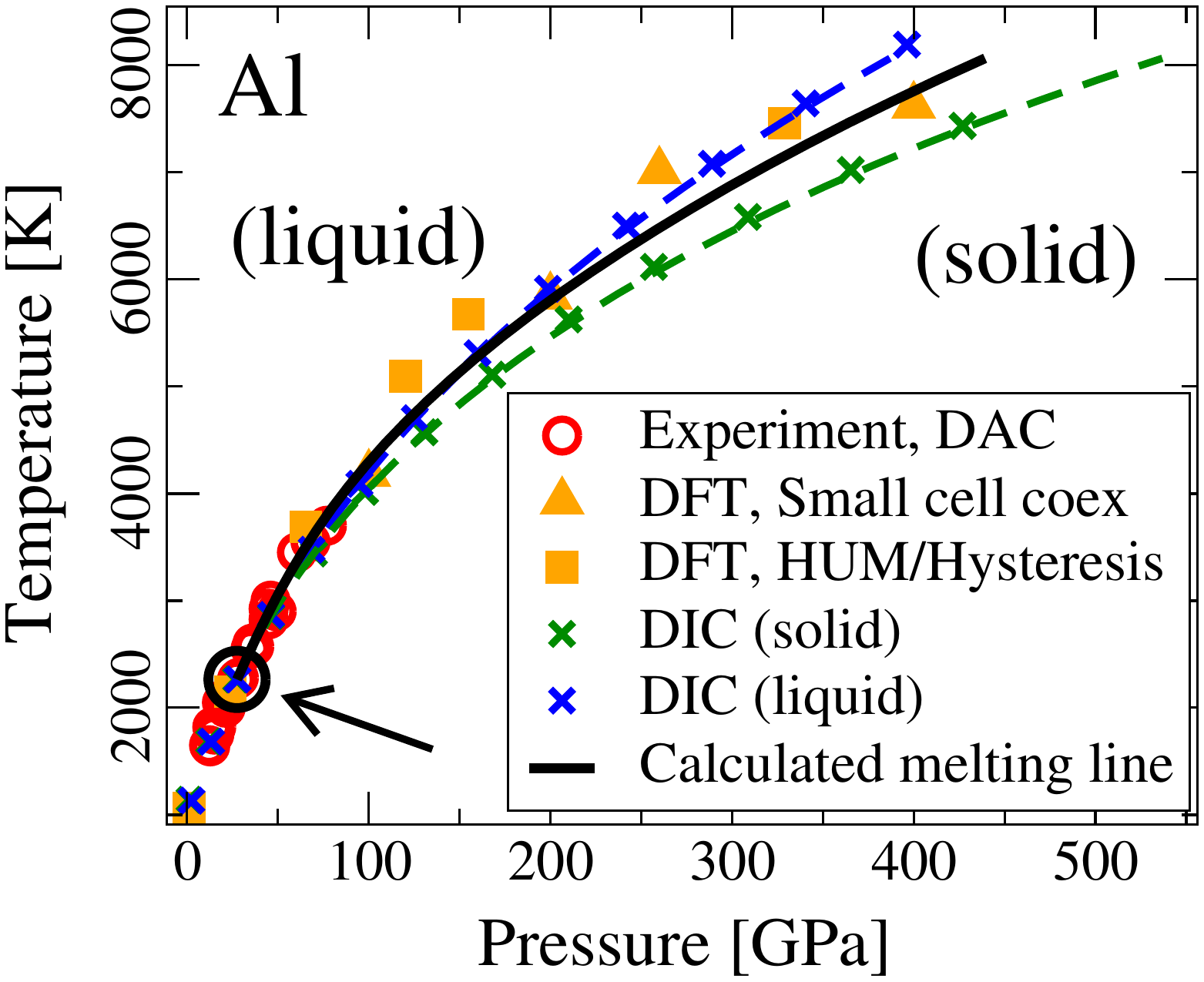}
    \caption{$(P,T)$ phase diagram of Al. The experimental point marked the black circle indicated by the arrow was used as reference point for tracing out the isomorphs. The predicted melting pressure is given by the solid black line. The experimental points are taken from Ref. \onlinecite{Hanstrom/Lazor:2000} which includes points from Ref. \onlinecite{Boehler/Ross:1997}, both determining melting in a diamond anvil cell (DAC). The DFT points are taken from Refs. \onlinecite{Hong/Walle:2019} (small cell coexistence method) and \onlinecite{Bouchet/others:2009} (HUM/Hysteresis).}
    \label{fig:Al_meltingdata}
\end{figure}

The resulting $(P,T)$ phase diagram is shown in Fig.~\ref{fig:Al_meltingdata}. The phase diagram also includes coexistence state points determined by experimental methods \cite{Hanstrom/Lazor:2000,Boehler/Ross:1997} (red, open circles) and by other simulation techniques using DFT \cite{Bouchet/others:2009,Hong/Walle:2019} (orange symbols). The DFT results depicted by the orange
triangles in Fig~\ref{fig:Al_meltingdata}
are obtained from PBE calculations.

\begin{figure}
    \centering
    \includegraphics[width=0.8\linewidth]{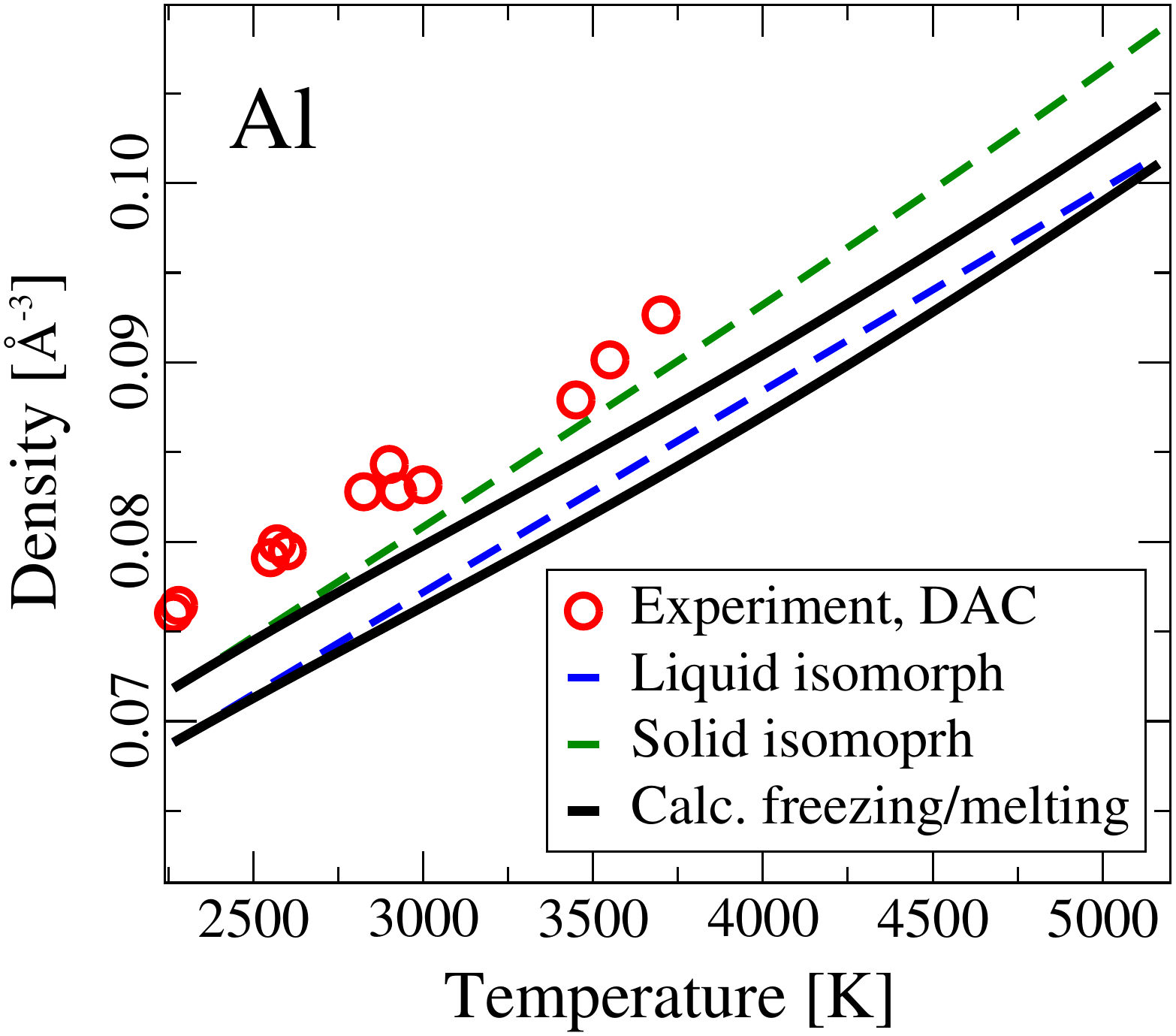}
    \caption{$(\rho,T)$ phase diagram of Al. Note that the axes are switched compared to the usual textbook way of plotting $(\rho,T)$ diagrams. This is because the method calculates densities as a function of temperature. The calculated melting pressure shown in Fig.~\ref{fig:Al_meltingdata} can be translated into the densities at which the liquid freezes and the solid melts via \eqref{eq:meltingrho}. The resulting densities are given by the solid black lines. The dashed lines mark the associated isomorphs. Reference \onlinecite{Hanstrom/Lazor:2000} reported only a single volume for each temperature, not stating for which phase. They are shown as densities here (red circles).}
    \label{fig:Al_densities}
\end{figure}

The reference point for the isomorphs is marked by the black circle (indicated by the arrow). The points along the isomorphs found using the DIC are marked by the colored crosses with the dashed line given by the polynomial fitted to the data. The DIC state points are determined by scaling the initial density by the factors listed in Sec.~\ref{sec:DFT-isomorphs}.

The solid isomorph lies to the right of the liquid in the $(P,T)$ phase diagram, that is, the solid isomorph is at  at a higher pressure than the liquid at the same temperature. This was seen also in the EMT-Cu results in Fig.~\ref{fig:EMT_results} (note that there pressure is the vertical axis), and can be rationalized by the following argument: The solid has a larger density and correspondingly a lower $\gamma$ than the liquid at the same temperature and pressure. For a given common temperature increase the solid therefore must undergo a larger fractional density increase. Assuming the bulk modulus of the solid is at least as large as that of the liquid, this implies a greater pressure increase.

The melting curve calculated from the two isomorphs is given by the solid black line in Fig.~\ref{fig:Al_meltingdata}. The calculated melting pressure agrees well with both the experimental and the DFT points. The latter, from Refs.~\onlinecite{Bouchet/others:2009} and \onlinecite{Hong/Walle:2019}, refer to the fcc crystal phase. In fact,
the liquid-fcc melting line continues until the liquid-fcc-bcc triple
point, predicted at 5650\,K and 195\,GPa by ab-initio DFT calculations.%
\cite{MultiphaseAlumSjostr2016}
Along the liquid-fcc melting line the deviation between
Ref.~\citenum{MultiphaseAlumSjostr2016}
and this work is less than 5\%.
Our predictions are closer to the data of Ref.~\onlinecite{Hong/Walle:2019} than to the experimental data, which is perhaps not surprising since the former were also based on DFT using the same functional as we did, and indeed the same software. The other DFT points from Ref.~\onlinecite{Bouchet/others:2009} are based on the hysteresis method, which combines data from the \textit{heat until it melts} (HUM) method with the opposite method, \textit{cool until it freezes}, to compensate for well-known overestimation by the the HUM method alone due to super-heating effects.\cite{Chokappa/Clancy:1987a,Chokappa/Clancy:1987b} The resulting combined melting curve is much better fit than HUM alone, but often still somewhat too high.\cite{Bouchet/others:2009}

Figure~\ref{fig:Al_densities} shows the densities calculated using \eqref{eq:meltingrho} together with the densities along the reference isomorphs. Unfortunately, there are no densities reported corresponding to the pressures
in the DFT calculations of Refs.~\onlinecite{Bouchet/others:2009} and \onlinecite{Hong/Walle:2019}, shown in Fig.~\ref{fig:Al_meltingdata}. Ref.~\onlinecite{Hanstrom/Lazor:2000} reports only a single volume for each temperature, which was translated to a density to obtain the red circles shown in Fig.~\ref{fig:Al_densities} (it is not clear from Ref.~\onlinecite{Hanstrom/Lazor:2000} whether the reported volume is for the liquid or the solid phase, but from Fig.~\ref{fig:Al_densities}, we presume that they correspond to the solid phase).

\section{\label{sec:reentrant_melting}Implications for the fate of the melting curve at high pressures}

In this section we discuss a potential consequence of the fact that the melting curve approximately follows an isomorph whose slope, given essentially by the density-scaling exponent $\gamma$, decreases substantially with increasing density. The question of how low $\gamma$ can become in the high-density limit encourages us to speculate about what isomorph theory can tell about the fate of the melting curve at very high pressures. In particular, inspired by recent results of Hong and van de Walle\cite{Hong/Walle:2019} we discuss when the the phenomenon known as {\em re-entrant melting} might occur.

When the melting temperature as a function of pressure has a maximum, then increasing pressure at a fixed temperature lower than the maximum causes the material to undergo a sequences of phases liquid$\rightarrow$crystal$\rightarrow$liquid, i.e., the system \emph{re-enters} the liquid phase. This phenomenon has been speculated on since the start of the last century\cite{Tammann:1903} and has come to be termed re-entrant melting. Re-entrant phenomena have been much studied in liquid crystals\cite{Mazza2011} and among metals; sodium is one case known to undergo re-entrant melting, this having been observed experimentally in 2005 \cite{Gregoryanz/others:2005} and later also in simulation.\cite{Hernandes/Iniguez:2007, Eshet2012} Re-entrant melting has also been seen in simple model systems such as that based on the Gaussian core pair potential.\cite{Stillinger1976,Lang2000}

\subsection{\label{sec:approach-Hong-vdW}The approach of Hong and van de Walle}

Hong and van de Walle \cite{Hong/Walle:2019} have recently suggested re-entrant melting to be far more common among metals than previously thought. The reason for this not being previously recognized is simply that the temperature maximum occurs at much 
higher pressures than could be studied in the laboratory so far. Based on DFT simulations of coexistence, however, they find re-entrant melting not just in Na but also in other metals, though typically at pressures well above what can be achieved in experiments. To locate the re-entrant point, they suggest a quick method to screen materials, thus avoiding the computationally demanding work of accurately determining the whole melting curve. The method is surprisingly simple and inadvertently related to isomorph theory as shown below.

According to the Clausius-Clapeyron relation, the slope of the melting curve is given by
\begin{equation}
    \frac{\text{d} P}{\text{d} T_m} = \frac{\Delta H}{T_m \Delta V}
\end{equation}
in which $\Delta H$ is the specific heat of fusion and $\Delta V = V_l - V_s$ is the difference in specific volume. The melting slope becomes negative whenever $\Delta V$ does, i.e., when $V_s > V_l$. The screening method of Hong and van de Walle relies on finding the pressure at which $V_s$ becomes larger than $V_l$, by taking a randomly selected snapshot from the trajectory of simulations of a solid and one from a liquid simulation. Each snapshot represents the corresponding phase. The snapshots are then compressed by scaling all atomic position vectors uniformly. From this the pressure-volume relation is estimated for each phase, and the pressure where the sign of $\Delta V$ changes can be found from comparison. The method provides a well-informed guess of where a melting curve maximum can be expected, if at all.  Hong and van de Walle\cite{Hong/Walle:2019} confirmed by simulations of Na, Mg  and Al that the maximum of the melting temperature $T_m$ coincides with the estimate from the screening method. For the case of Al, the re-entrant point was located at the rather extreme conditions of around 3500\,GPa and 20000\,K. 

The technique of uniformly scaling configurations to estimate high-pressure thermodynamic properties is very similar in spirit to isomorph theory, in particular the above used direct isomorph check (DIC), and can be said to implicitly assume that the freezing and melting lines are isomorphs of the respective phases. Scaling a single configuration can only give a rough estimate of the pressure, not of the temperature at the higher density, while the DIC, by sampling several configurations, enables an estimate of the corresponding temperature as well. Note, however, that a recent paper described a force-based method for tracing out an isomorph based on a single configuration.\cite{Schroder:2022}

\subsection{\label{sec:isomorph-Lindemann-approach}Approach based on isomorph theory and the Lindemann criterion}

A possible method for a similarly educated guess derived from isomorph theory directly could be based on the density-scaling exponent $\gamma$. The connection between $\gamma$ going to zero and the maximum of the melting is more than a hand-waving argument of the slope flattening out. In the Lindemann law \cite{Lindemann:1910,Gilvarry:1956} for melting, the melting curve is described by the function
\begin{align}
    \frac{\mathrm{d ln} T_m}{\mathrm{d} P} = \frac{2 \left( \gamma_{G,m} - \frac{1}{3} \right) }{K_{T,m}},
    \label{eq:Lindemann}
\end{align}
where $K_{T,m}$ is the isothermal bulk modulus of the solid at melting and $\gamma_{G,m}=\alpha_{P,m}K_{T,m}V_m/\rho c_{V,m}$ is the thermodynamic Gr\"uneisen parameter at melting, here given in terms of $K_T$, the isobaric thermal expansion coefficient $\alpha_P$, and the isochoric heat capacity per particle $c_V$. The Lindemann law is often used to extrapolate the melting curve to high pressures from low pressure data. Typical values for $\gamma_{G}$ in metals at low pressure are around $2$ and higher,\cite{Burakovsky/Preston:2004} thus yielding a melting curve with a positive slope. Writing this law in terms of the density rather than pressure gives
\begin{equation}
     \frac{\mathrm{d ln} T_m}{\mathrm{d} \ln\rho} = 2 \left( \gamma_{G,m} - \frac{1}{3} \right).
    \label{eq:Lindemann_rho}
\end{equation}
A simple relation between the Gr\"uneisen parameter $\gamma_G$ and the scaling exponent $\gamma$ from isomorph theory was derived some time ago for spherical particles.\cite{paper3,Hummel/others:2015} Using the Dulong-Petit approximation for the specific heat, the relation is 
\begin{align}
    \gamma = 2 \left( \gamma_G - \frac{1}{3} \right)\,.
    \label{eq:gamma_to_grueneisen}
\end{align}
Using this the Lindemann melting law becomes simply
\begin{equation}
     \frac{\mathrm{d ln} T_m}{\mathrm{d} \ln\rho} = \gamma(\rho),
    \label{eq:Lindemann_rho_gamma}
\end{equation}
which is identical to the equation for an isomorph, \eqref{eq:gamma_diff}. In writing this equation we have removed the subscript $m$ referring explicitly to the melting line from $\gamma$ and assumed that it depends only on $\rho$. Thus under the Dulong-Petit approximation the Lindemann melting law is equivalent to the melting line following a (crystal) isomorph. From Eqs. (\ref{eq:Lindemann}) and (\ref{eq:Lindemann_rho_gamma}) it follows that the slope of the melting curve becomes negative when $\gamma_G < 1/3$ or equivalently $\gamma<0$. This cannot be taken as a strict criterion, however, because (1) the melting line does not strictly follow an isomorph and (2) using the isomorph as the basis for a perturbative approach as we have done is unlikely to work as $\gamma$ approaches zero; this is because the virial potential-energy correlation coefficient $R$ goes to zero when $\gamma$ does, meaning that isomorph theory is not expected to work in this region. 


What is the physical origin of re-entrant melting in metals? The authors of Ref.~\onlinecite{Hong/Walle:2019} explain the origin of re-entrant melting as a faster softening of interatomic interactions in the liquid phase than in the solid. The decrease of $\gamma$ also corresponds to a softening of interactions in the sense of an effective IPL\cite{Hummel/others:2015} whose exponent decreases -- the interactions become effectively more long ranged. Indeed, a value of zero for $\gamma$ formally implies interactions independent of distance, which would imply that there is no energy cost to melting: Increasing pressure of the coexisting phases normally requires an increase in temperature so that the entropic term in the free energy can continue to balance the enthalpic term, assuming the entropy of fusion does not change much. If the energy cost of melting vanishes upon increasing pressure, then this is no longer necessary and no increase in temperature is needed to maintain coexistence.

The softening of the effective interatomic potential can give rise to other complications for our isomorph-based method, in particular a change of crystal phase, since smaller IPL exponents typically favor more open crystal structures such as bcc.\cite{Hummel/others:2015} 
In fact, Na, the metallic case for which re-entrant melting has been observed experimentally, has a bcc structure under ambient conditions; this is related to sodium's fairly low value of $\gamma\sim1.9$ (liquid phase near the triple point\cite{Hummel/others:2015}), meaning presumably that lower pressures are required to reach $\gamma=0$. Simulations on Na show that the correlation coefficient $R$ is also low in the relevant range\cite{Friedeheim:2020}, and consequently the isomorphs are of poor quality in regard to having approximately invariant structure and dynamics. Nevertheless, the isomorph-predicted melting curve is close to existing DFT results and largely consistent with the experimental curve.\cite{Friedeheim:2020} For materials such as Al and Cu, which have larger $\gamma$ and close-packed structures at ambient pressure, a phase change to bcc\cite{MultiphaseAlumSjostr2016, Hong/Walle:2019} is indicated by the lowering of $\gamma$ and the consequent softening of effective interactions with increasing density. Thus, structure is inherently not invariant beyond the phase change, precluding an accurate isomoprh-based prediction of the melting line all the way up to $\gamma\rightarrow0$ starting from the close-packed phase. Still, locating the vanishing of $\gamma$ ignoring a possible bcc transition could be an indicator of the possibility of re-entrant melting. We note that the fcc-bcc phase transition for Al has recently been identified at pressures at around 200\,GPa.\cite{MultiphaseAlumSjostr2016}

\subsection{\label{sec:comparison-EXP}Insights from a comparison with the EXP pair-potential system}

 To shed further light on the possibility of re-entrant melting, we consider a simple model system which exhibits similar behavior to metals regarding the density dependence of $\gamma$. This is the purely repulsive pair potential given by a single decaying exponential, denoted EXP: 

\begin{equation}
    v_\text{EXP}(r)= \varepsilon\, \text{e}^{-r/\sigma}
\end{equation}
The EXP pair-potential model describes certain aspects of metals surprisingly well. This can be traced back to the fact that the low-density limit of the Yukawa (screened Coulomb) potential, an important ingredient in most models for metals,\cite{Nikoofard/Hajiashrafi:2016} is well described by the EXP potential.\cite{Bacher/others:2020} The connection can also be rationalized by noticing that the EMT description of metals involves an exponential function in several places.\cite{Jacobsen/others:1996} What is particularly relevant in the present context is that the density dependence of $\gamma$ is similar to what is observed for metals, exhibiting substantial decay, in fact vanishing at sufficiently high density.\cite{Pedersen/Bacher/others:2019} This density dependence is exhibited both by EMT and DFT metals. The EMT description contains the most important aspects of the quantum-mechanical description of metals at high densities, while at the same time having an analytical connection to the simple EXP system. An analytic relation between $\gamma$ and the parameters of the EMT potential has not yet been found, but the latter presumably inherits its density dependence from the EXP pair-potential system, which is thus somehow the root of the behavior in DFT (and presumably real) metals.

Considering the EXP system as a prototype for metals is consistent with the hypothesis of Ref. \onlinecite{Hong/Walle:2019} that re-entrant melting applies for all metals; the melting curve for the EXP potential indeed has a maximum temperature.\cite{Pedersen/Bacher/others:2019} In fact, $\gamma_\text{EXP}$ goes to zero almost exactly where the maximum of $T_m$ is reached. \cite{Bacher/others:2020} A criterion based on fitting and interpolating the zero crossing of $\gamma$ could therefore be used similarly to the volume screening method to outline the region where the maximum of the melting curve should be expected. Note, however that we do not claim one can extrapolate using isomorphs to accurately locate the re-entrant point. In fact, the theory breaks down as $\gamma$ vanishes. A possible advantage over the volume screening method would be that using the $\gamma$ interpolation method can be done from scaling of only one phase rather than both -- although it uses information from more than one configuration. 

Decreasing $\gamma$ with increasing density and the appearance of the bcc phase at high pressures are both consistent with metallic interactions being well-described by the EXP pair potential.
The EXP system also undergoes an fcc-bcc phase transition and eventually reaches re-entrant melting.\cite{Pedersen/Bacher/others:2019} Conversely, if the EXP system is indeed a good model for metallic interactions,
the proposition of Hong and van de Walle follows that re-entrant melting is widespread among metals, possibly universal.

\subsection{\label{sec:other-approaches}Other approaches}

We round off this section with some discussion of related approaches. Clearly, knowing the density dependence of either the density scaling exponent $\gamma$ or the Gruneisen parameter $\gamma_G$ is relevant to determining the melting curve. In the literature the Gruneisen parameter is the more commonly studied quantity. For example Coung and Phan\cite{Cuong2021} have used analytical expressions based on atomistic modelling and experimental data to estimate the melting curve of iron up to 350 GPa. An important ingredient of their approach is an expression for the density-dependence of the Gruneisen parameter. They used a simple power-law form, similar to what we have used in our consistency check of the DIC, though they note that other forms are also found in the literature, for example a constant plus two power laws.\cite{Hoc2020} Another recent example of work connecting the density dependence of the Gruneisen parameter to melting curves for metals is by Roy and Sarker.\cite{Roy/Sarkar:2021} These authors find using DFT calculations of the crystal in the harmonic approximation that for Pd, Rh, Pt and Ir, both the vibrational and thermal Gruneisen parameters exhibit a density dependence described by

\begin{equation}
    \frac{\rho}{\gamma_G} = a + \rho\,.
\end{equation}
That is, the ratio of density to the Gruneisen parameter is a linear function of density with slope unity. Their data covers large density changes corresponding to pressures up to 350 GPa and the quality of their linear fits is excellent. This result (at least for the vibrational version of $\gamma_G$) is equivalent to the Debye frequency also being a linear function of density. It also implies, though the authors do not discuss it, that $\gamma_G$ levels out at value unity in their high-density limit, which by \eqref{eq:gamma_to_grueneisen} corresponds to a limiting value of $\gamma\rightarrow4/3$. They note that experimental data seems to give a somewhat higher slope, around 1.3, corresponding to a limiting value $\gamma_G \rightarrow \sim 0.77$ and $\gamma\rightarrow \sim 0.87$.  Our data cannot be used to infer a limiting value of $\gamma$, but it appears that it will quite likely continue to values lower than 4/3 at least (Fig.~\ref{fig:Al_DIC_check}). Roy and Sarkar's result of a limiting value of $\gamma_G$ greater than 1/3 precludes the possibility of re-entrant melting, assuming that \eqref{eq:Lindemann} describes the melting curve. Therefore their results are at odds with those of Ref.~\onlinecite{Hong/Walle:2019}. It would be interesting to attempt to fit their data with functional forms which do not preclude re-entrant melting; this could potentially reconcile their data with 
those of Hong and van de Walle.

To summarize this section, the work of Hong and van de Walle\cite{Hong/Walle:2019} has shown that computational studies of melting can provide new and unexpected insights. Our results supplement this with insight from the study of isomorphs in both realistic and model systems. They show the utility of the isomorph approach as a practical tool for determining melting curves and support the conjecture of Hong and van de Walle that re-entrant melting is possibly universal among metals.



\section{Summary}

We have studied numerically an EMT model of Cu and carried out DFT simulations of Al with a focus on freezing and melting. Isomorph-theory based predictions were largely confirmed, demonstrating the possibility of estimating the pressure and temperature variation at melting from a single reference state point. Our findings are consistent with the conjecture of Hong and van de Walle of possibly universal re-entrant melting at extremely high pressures.

\begin{acknowledgments}
The work was supported by the VILLUM Foundation's {\em Matter} Grant (No. 16515). The computational results presented have been achieved in part using the Vienna Scientific Cluster (VSC).
\end{acknowledgments}

\appendix

\section{\label{app:thermodynamic_data}Isomorph data}

Tables~\ref{tab:Cu_solid}, \ref{tab:Cu_liquid}, ~\ref{tab:Al_solid} and \ref{tab:Al_liquid} give thermodynamic data along the isomorphs for all the isomorphs studied, both solid and liquid for Cu and Al.

\begin{table}
    \centering
    \begin{tabular}{ c c c | c c }
        $T$ [K] & $P$ [GPa] & $\rho$ [$\text{\AA}^{-3}$]~ & ~~ $ ~ \gamma ~ $ ~~ & ~~ $ ~ R ~ $ ~~ \\[1mm] \hline
 2008 & 16.0 & 0.0873 & 3.160 & 0.9906 \\ 
 2168 & 20.6 & 0.0894 & 3.008 & 0.9915 \\ 
 2331 & 25.6 & 0.0917 & 2.876 & 0.9921 \\ 
 2499 & 31.1 & 0.0940 & 2.760 & 0.9927 \\ 
 2672 & 37.0 & 0.0963 & 2.657 & 0.9932 \\ 
 2850 & 43.4 & 0.0987 & 2.558 & 0.9935 \\ 
 3033 & 50.4 & 0.1012 & 2.469 & 0.9939 \\ 
 3221 & 57.9 & 0.1037 & 2.384 & 0.9939 \\ 
 3413 & 66.0 & 0.1063 & 2.303 & 0.9941 \\ 
 3609 & 74.8 & 0.1090 & 2.224 & 0.9942 \\ 
 3809 & 84.2 & 0.1117 & 2.148 & 0.9942
    \end{tabular}
    \caption{Temperature, pressure, density, as well as $\gamma$ and $R$ of the state points found from the DIC for solid Cu. 
    }
    \label{tab:Cu_solid}
\end{table}

\begin{table}
    \centering
    \begin{tabular}{ c c c | c c }
        $T$ [K] & $P$ [GPa] & $\rho$ [$\text{\AA}^{-3}$]~ & ~~ $ ~ \gamma ~ $ ~~ & ~~ $ ~ R ~ $ ~~ \\[1mm] \hline
 2008 & 16.0 & 0.0831 & 3.378 & 0.9730 \\ 
 2179 & 20.3 & 0.0852 & 3.210 & 0.9760 \\ 
 2356 & 24.9 & 0.0873 & 3.058 & 0.9786 \\ 
 2536 & 30.0 & 0.0895 & 2.925 & 0.9804 \\ 
 2723 & 35.5 & 0.0917 & 2.799 & 0.9817 \\ 
 2914 & 41.4 & 0.0940 & 2.690 & 0.9835 \\ 
 3111 & 47.9 & 0.0964 & 2.586 & 0.9843 \\ 
 3312 & 54.9 & 0.0988 & 2.489 & 0.9852 \\ 
 3519 & 62.4 & 0.1012 & 2.398 & 0.9858 \\ 
 3730 & 70.6 & 0.1038 & 2.316 & 0.9866 \\ 
 3946 & 79.4 & 0.1064 & 2.237 & 0.9870
    \end{tabular}
    \caption{Temperature, pressure, density, as well as $\gamma$ and $R$ of the state points found from the DIC for liquid Cu. 
    }
    \label{tab:Cu_liquid}
\end{table}

\begin{table}
    \centering
    \begin{tabular}{c | c c c | c c }
        $\rho/\rho_0$ & $T$ [K] & $P$ [GPa] & $\rho$ [$\text{\AA}^{-3}$] & $\gamma$ & $R$ \\[1mm] \hline
0.8 & 1141 & 1.35 & 0.0575 & 3.552 & 0.9759 \\
0.9 & 1687 & 12.15 & 0.0647 & 2.998 & 0.9849 \\
1.0 & 2265 & 27.07 & 0.0719 & 2.591 & 0.9896 \\
1.1 & 2854 & 46.27 & 0.0790 & 2.281 & 0.9920 \\
1.2 & 3441 & 69.87 & 0.0862 & 2.041 & 0.9928 \\
1.3 & 4015 & 97.96 & 0.0934 & 1.851 & 0.9923 \\
1.4 & 4573 & 130.64 & 0.1006 & 1.696 & 0.9906 \\
1.5 & 5111 & 167.97 & 0.1078 & 1.568 & 0.9877 \\
1.6 & 5626 & 210.05 & 0.1150 & 1.458 & 0.9834 \\
1.7 & 6117 & 256.93 & 0.1221 & 1.362 & 0.9775 \\
1.8 & 6583 & 308.68 & 0.1293 & 1.276 & 0.9697 \\
1.9 & 7022 & 365.34 & 0.1365 & 1.198 & 0.9596 \\
2.0 & 7433 & 426.95 & 0.1437 & 1.125 & 0.9465
    \end{tabular}
    \caption{Temperature, pressure, density, as well as $\gamma$ and $R$ of the state points found from the DIC for solid Al. 
    }
    \label{tab:Al_solid}
\end{table}

\begin{table}
  \begin{tabular}{c | c c c | c c }
        $\rho/\rho_0$ & $T$ [K] & $P$ [GPa] & $\rho$ [$\text{\AA}^{-3}$] & $\gamma$ & $R$ \\[1mm] \hline
0.8 & 1123 & 2.26 & 0.0550 & 3.562 & 0.9393 \\
0.9 & 1680 & 13.03 & 0.0619 & 3.009 & 0.9632 \\
1.0 & 2265 & 27.82 & 0.0688 & 2.616 & 0.9679 \\
1.1 & 2865 & 46.23 & 0.0757 & 2.361 & 0.9740 \\
1.2 & 3476 & 68.57 & 0.0826 & 2.165 & 0.9772 \\
1.3 & 4090 & 94.92 & 0.0894 & 2.010 & 0.9786 \\
1.4 & 4704 & 125.35 & 0.0963 & 1.883 & 0.9787 \\
1.5 & 5312 & 159.93 & 0.1032 & 1.777 & 0.9777 \\
1.6 & 5912 & 198.69 & 0.1101 & 1.687 & 0.9760 \\
1.7 & 6502 & 241.69 & 0.1170 & 1.609 & 0.9734 \\
1.8 & 7080 & 288.94 & 0.1238 & 1.541 & 0.9702 \\
1.9 & 7644 & 340.48 & 0.1307 & 1.480 & 0.9663 \\
2.0 & 8193 & 396.32 & 0.1376 & 1.425 & 0.9618
    \end{tabular}
    \caption{Temperature, pressure, density, as well as $\gamma$ and $R$ of the state points found from the DIC for liquid Al. 
    }
    \label{tab:Al_liquid}
\end{table}

\section{\label{app:pressure_correction}Correction to the pressure}

The pressure and temperature of solid/liquid coexistence from interface pinning are determined in $NPT$ simulations. However, in order to generate isomorphs simulations need to be carried out in the $NVT$ ensemble. This means the pressure is no longer a fixed quantity but one that experiences statistical fluctuations, causing the following issue for determining the melting pressure via Eq.~(\ref{eq:meltingP}): The starting point for the reference isomorphs has been previously determined from interface pinning (or any other method of choice) and should be at coexistence, meaning that $P_s^0=P_l^0=P_m^0$. Setting $T=T^0$ in Eq.~(\ref{eq:meltingP}) 
and then using $P V = N k_B T + W$ we find
\begin{align}
    P_m(T=T^0) &= \frac{ W_l^0 - W_s^0 }{V_l^0 - V_s^0} = \frac{ P_l^0 V_l^0 - P_s^0 V_s^0 }{ V_l^0 - V_s^0 } \nonumber
    \\
    &= P_m^0 \frac{ V_l^0 - V_s^0 }{ V_l^0 - V_s^0 } = P_m^0 \label{eq:PressureProblem}
\end{align}

in the case of the simulated pressure being exactly the pressure found for coexistence. In reality, however, the pressures on both the liquid and solid  sides are subject to statistical fluctuations. Assuming small deviations, $\sigma$, from the expected melting pressure $P_{m,exp}^0$ for $P_s^0$ and $P_l^0$, respectively, Eq.~(\ref{eq:PressureProblem}) instead becomes 
\begin{align}
    P_m &= \frac{\left( (P_{m,exp}^0 \pm \sigma_{P,l}^0) V_l^0 - (P_{m,exp}^0 \pm \sigma_{P,s}^0) V_s^0 \right)}{\left(V_l^0 - V_s^0\right)}\nonumber
    \\
    &= P_{m,exp} \pm \frac{ \sigma_{P,l}^0 V_l^0}{V_l^0 - V_s^0} \pm \frac{ \sigma_{P,s}^0 V_-^0}{V_l^0 - V_s^0}\label{eq:PressureProblem2}
\end{align}
where the ratio between the volume of a phase $V_{s/l}$ versus the difference between the two phases gives a magnifying factor of $\sim10$. This makes the contribution from the initially small deviations $\sigma_{P,s}^0$ and $\sigma_{P,l}^0$ large. 

To obtain sensible results we apply an heuristic approach to correcting this error. It is a minimal correction which guarantees that the consistency check of \eqref{eq:PressureProblem} is satisfied. First, we assume that the errors in the volumes corresponding to the reference melting pressure are small enough to be neglected. This is justified since the pressure-volume relations for the solid and liquid phases were determined not from single simulations but from fitting the pressure for a series of $NVT$ simulations. Thus, taking $P_s^0$ and $P_l^0$ to be the value obtained from interface pinning at the reference point, $P_m^0$, and using $P V = N k_B T + W$, we know the ``true'' average values of the virials for both phases (note that although pressure and temperature are the same, the volumes differ and therefore so do the virials). The difference between the true and measured virials, which defines the correction to the virial at the reference point $W^0_{NVT}$, is given by

\begin{align}
    \Delta W = W_{l,s;m} - W^0_{l,s;NVT} = (P^0_{m} - P_{l,s}^0) V_{l,s}^0
\end{align}
where all quantities on the right side are to be understood as thermodynamic averages. This correction to the reference-point virials in \eqref{eq:meltingP} guarantees that the consistency check is satisfied.

\end{document}